\documentclass{SIP}

\usepackage{epstopdf, epsfig, subcaption}

\begin{document}

\title{Recent Advances in End-to-End Automatic Speech Recognition}

\author{Jinyu Li}

\address{
Address: 1 Microsoft Way, Redmond, WA, 98052, USA
}

\corres{\name{Jinyu Li}
\email{jinyli@microsoft.com}}

\begin{abstract}
Recently, the speech community is seeing a significant trend of moving from deep neural network based hybrid modeling to end-to-end (E2E) modeling for automatic speech recognition (ASR). While E2E models  achieve the state-of-the-art results in most benchmarks in terms of ASR accuracy, hybrid models are still used in a large proportion of  commercial ASR systems at the current time. There are lots of practical factors that affect the production model deployment decision. Traditional hybrid models, being optimized for production for decades, are usually good at these factors. Without providing excellent solutions to all these factors, it is hard for E2E models to be widely commercialized. In this paper, we will overview the recent advances in E2E models, focusing on technologies addressing those challenges from the industry's perspective. 
\end{abstract}


\maketitle

\section{Introduction}

The accuracy of automatic speech recognition (ASR) has been significantly boosted since deep neural network (DNN) based hybrid modeling \cite{DNN4ASR-hinton2012} was adopted a decade ago. This breakthrough used DNN to replace the traditional Gaussian mixture model for the acoustic likelihood evaluation, while still keeping all the components such as acoustic model, language model, and lexicon model, etc.,  as the hybrid ASR system. Recently, the speech community has a new breakthrough by transiting from hybrid modeling to end-to-end (E2E) modeling \cite{graves2014towards, hannun2014deep, chorowski2014end, miao2015eesen, bahdanau2016end, chan2016listen, collobert2016wav2letter, prabhavalkar2017comparison} which directly translates an input speech sequence into an output token sequence using a single network. Such a breakthrough is even more revolutionary because it overthrows all the modeling components in traditional ASR systems, which have been used for decades. 

There are several major advantages of E2E models over traditional hybrid models. First, E2E models use a single objective function which is consistent with the ASR objective to optimize the whole network, while traditional hybrid models optimize individual components separately, which cannot guarantee the global optimum. Therefore, E2E models have been shown to outperform traditional hybrid models not only in academics \cite{watanabe2017hybrid} but also in the industry \cite{sainath2020streaming, Li2020Developing}.  Second, because E2E models directly output characters or even words, it greatly simplifies the ASR pipeline. In contrast, the design of traditional hybrid models is complicated, requiring lots of expert knowledge with years of ASR experience. Third, because a single network is used for ASR,  E2E models are much more compact than traditional hybrid models. Therefore, E2E models can be deployed to devices with high accuracy.

While E2E models  achieve the state-of-the-art results in most benchmarks in terms of ASR accuracy,    hybrid models are still used in a large proportion of  commercial ASR systems at the time of writing because the ASR accuracy is not the only factor for the production choice between hybrid and E2E models. There are lots of practical factors such as streaming, latency, adaptation capability, etc., which affect the commercial model deployment decision. Traditional hybrid models, optimized for production for decades, are usually good at these factors. Without providing excellent solutions to all these factors, it is hard for E2E models to be widely commercialized. To that end, in  this paper, we overview popular E2E models with a focus on the technologies addressing those challenges from the perspective of the industry.

\section{End-to-end models}
\label{sec:E2E}
\begin{figure}
     \centering
     \begin{subfigure}[b]{0.15\textwidth}
         \centering
         \includegraphics[width=\textwidth]{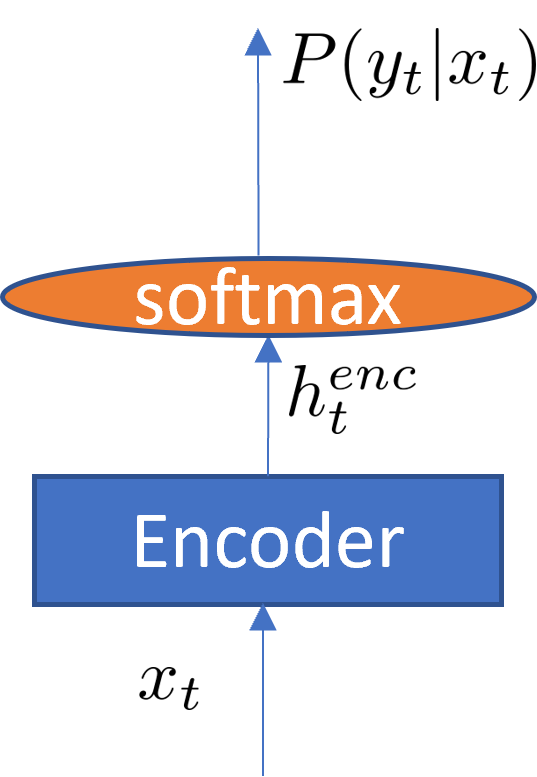}
         \caption{CTC}
         \label{fig:CTC}
     \end{subfigure}
     \hfill
     \begin{subfigure}[b]{0.25\textwidth}
         \centering
         \includegraphics[width=\textwidth]{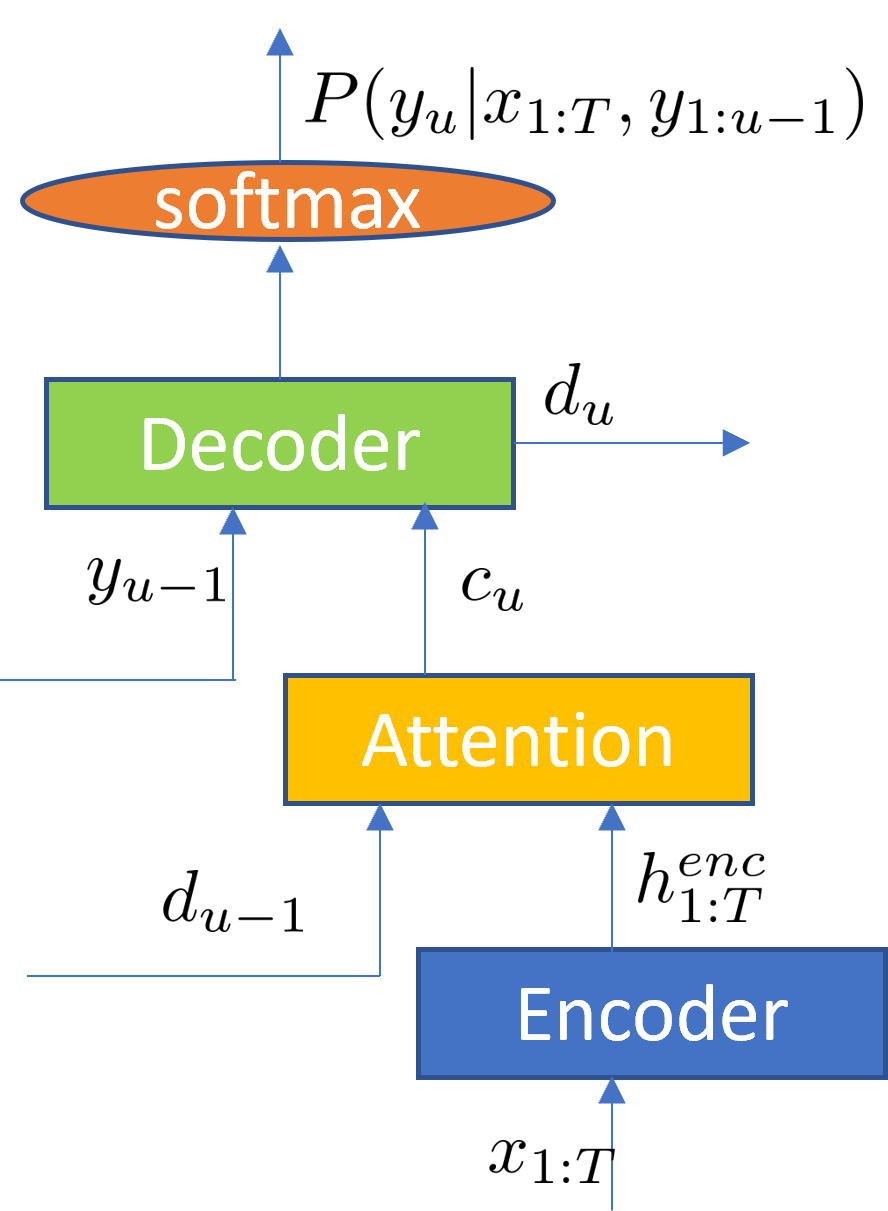}
         \caption{AED}
         \label{fig:AED}
     \end{subfigure}
     \hfill
     \begin{subfigure}[b]{0.3\textwidth}
         \centering
         \includegraphics[width=\textwidth]{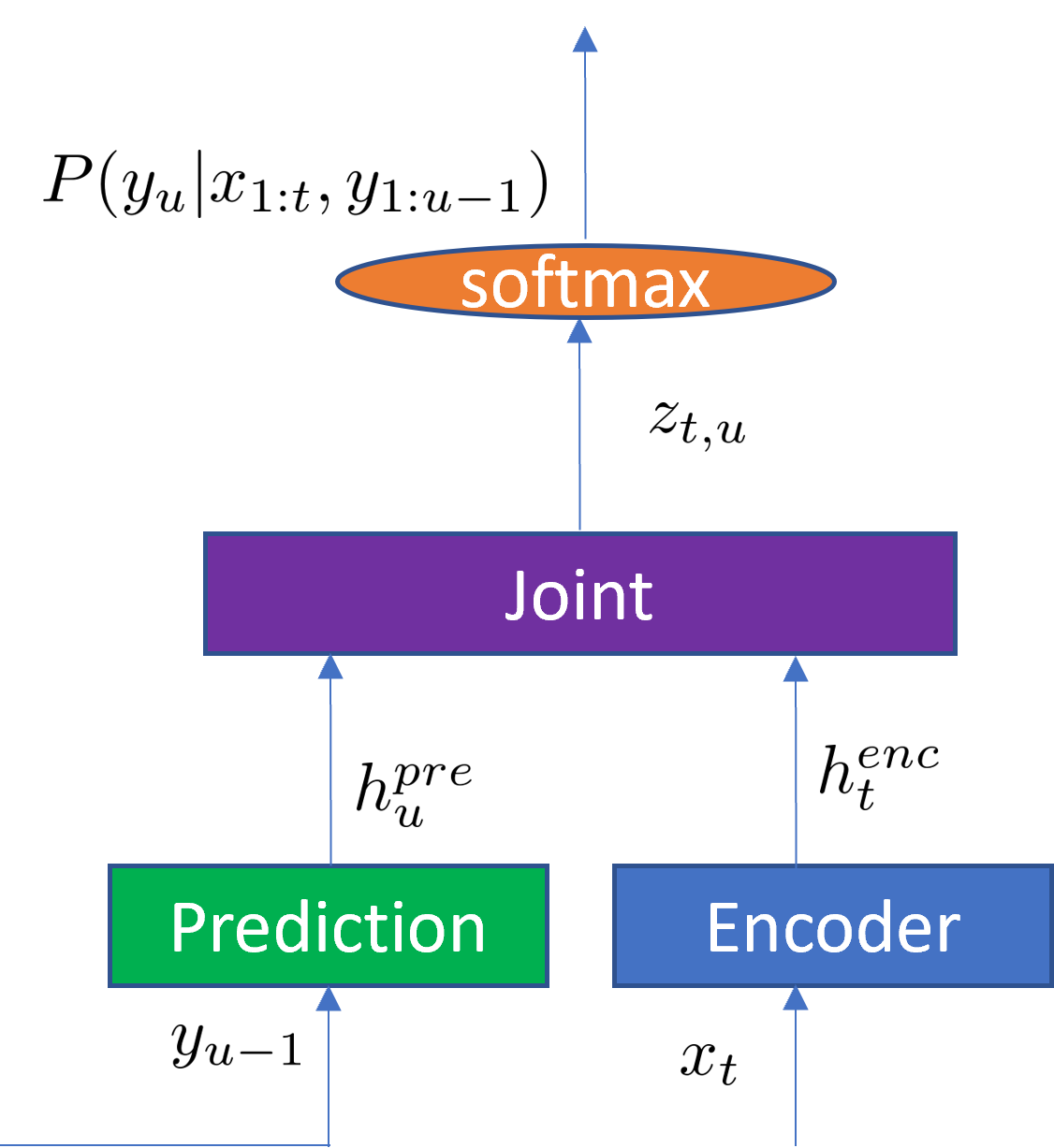}
         \caption{RNN-T}
         \label{fig:RNNT}
     \end{subfigure}
        \caption{Architectures of three popular end-to-end techniques \cite{he2019streaming}}
        \label{fig:e2e}
\end{figure}

The most popular E2E techniques for ASR are: (a) Connectionist Temporal Classification (CTC) \cite{graves2006connectionist}, (b) Attention-based Encoder-Decoder (AED) \cite{cho2014learning, Attention-bahdanau2014}, and (c) recurrent neural network Transducer (RNN-T) \cite{Graves-RNNSeqTransduction}. Among them, RNN-T provides a natural solution to streaming ASR with high accuracy and low latency, ideal for industrial application. 
These three most popular E2E techniques are illustrated in Figure \ref{fig:e2e}. In this section, we will give a short overview of them. 

\subsection{Connectionist Temporal Classification} \label{sec:ctc} 

\begin{figure}[t]
  \centering
  \includegraphics[width=0.9\linewidth]{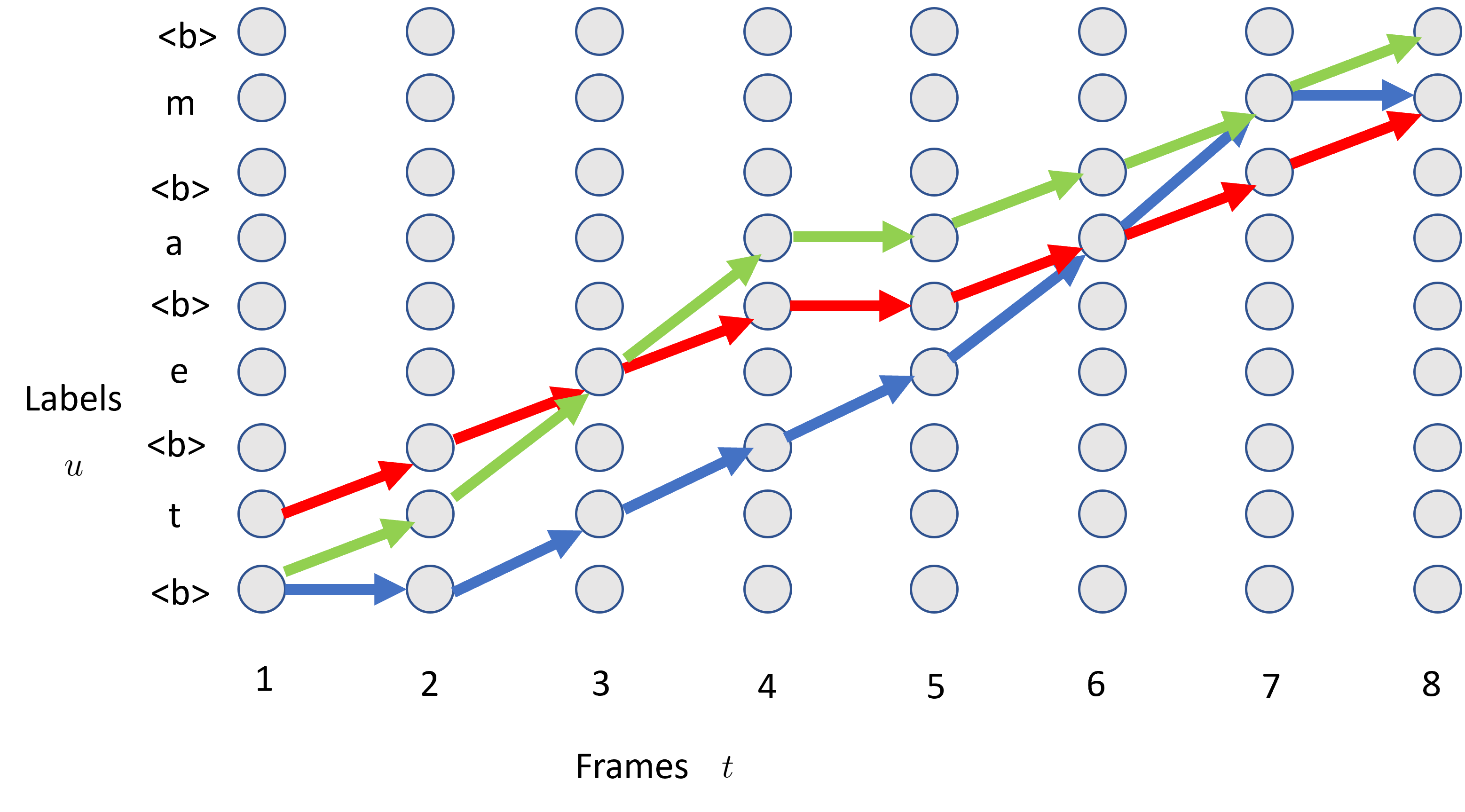}
  \caption{Example CTC paths for the word ``team''. A blank label $\langle b \rangle$  is inserted between every character. The blue path is ($\langle b \rangle$, $\langle b \rangle$ , t, $\langle b \rangle$ , e, a, m, m). The green path is ($\langle b \rangle$, t, e, a, a, $\langle b \rangle$, m, $\langle b \rangle$). The red path is (t, $\langle b \rangle$, e, $\langle b \rangle$, $\langle b \rangle$, a, $\langle b \rangle$, m).}
  \label{fig:CTC_path}
\end{figure}

The Connectionist Temporal Classification (CTC) technique for ASR \cite{graves2014towards} was designed to map the speech input sequence into an output label sequence. Because the length of output labels is smaller than that of the input speech sequence, a blank label is inserted between output labels with allowable repetition of labels to construct CTC paths that have the same length as the input speech sequence.  Figure \ref{fig:CTC_path} shows an example of three CTC paths for the word ``team''. 

Figure \ref{fig:CTC} shows the architecture of CTC. We denote the input speech  sequence as $\bf{x}$, the original output label sequence as $\bf{y}$, and  all of the CTC paths mapped from $\bf{y}$ as $B^{-1}(\bf{y})$. The encoder network is used to convert the acoustic feature $x_t$ into a high-level representation ${\bf{h}}_t^{enc}$. The CTC loss function is defined as the negative log probabilities of correct labels given the input speech sequence:
\begin{equation}
L_{CTC} = - ln P( \bf{y}|\bf{x} ),
\end{equation}
with 
\begin{equation} \label{eq:Prob_CTC}
P( {\bf{y}|\bf{x}} ) = \sum_{\bf{q} \in B^{-1}(\bf{y})} P( \bf{q} | \bf{x} ),
\end{equation}
where $\bf{q}$ is a CTC path. With the conditional independence assumption, $P( \bf{q} | \bf{x} )$ can be decomposed into a product of frame posterior as
\begin{equation}
P( {\bf{q} | \bf{x}} ) = \prod_{t=1}^T P( q_t | \bf{x}),	\label{eq:ctc_posterior}
\end{equation}
where $T$ is the length of the speech sequence. 

CTC is the first E2E technology widely used in ASR \cite{hannun2014deep, miao2015eesen, soltau2016neural, zweig2017advances, zeyer2017ctc, li2018advancing, audhkhasi2018building}. However, the conditional independence assumption in CTC is most criticized. One way to relax that assumption is to use the attention mechanism \cite{Das18advancing, salazar2019self} which can introduce implicit language modeling across speech frames.  Such attention-based CTC models implicitly relax the conditional independence assumption by improving the encoder without changing the CTC objective function, and therefore enjoy the simplicity of CTC modeling. By  replacing the underlying long short-term memory (LSTM) \cite{Hochreiter1997long}  with Transformer \cite{vaswani2017attention} in the encoder, which allows a more powerful attention mechanism to be used, CTC  thrives again in recent studies \cite{higuchi2020mask}. It gets further boosted by the emerged self-supervised learning technologies \cite{baevski2020wav2vec, chung2020generative, zhang2020pushing, hsu2021hubert} which can learn a very good representation that carries semantic information. 


\subsection{Attention-based Encoder-Decoder} \label{ssec:attention}

The attention-based encoder-decoder (AED) model is another type of E2E ASR model \cite{chorowski2014end, bahdanau2016end, chan2016listen, lu2016training, zeyer2018improved}. As shown in Figure \ref{fig:AED}, AED has an encoder network, an attention module, and a decoder network. 
The AED model calculates the probability as  
\begin{equation} \label{eq:Prob_attention} 
P( {\bf{y}|\bf{x}} ) = \prod_u P(y_u | {\bf{x}} , {\bf{y}}_{1:u-1}),
\end{equation}
where $u$ is the output label index. The training objective is also to minimize $- ln P( {\bf{y}}|{\bf{x}} )$. 

The encoder network performs the same function as the encoder network in CTC by converting input feature sequences into high-level hidden feature sequences. The attention module computes attention weights between the previous decoder output and the encoder output of each frame using attention functions such as additive attention \cite{bahdanau2016end} or dot-product attention \cite{chan2016listen}, and then generates a context vector as a weighted sum of the encoder outputs. The decoder network takes the previous output label together with the context vector to generate its output to calculate $P(y_u | {\bf{x}} , {\bf{y}}_{1:u-1})$, which is operated in an autoregressive way as a function of the previous label outputs without the conditional independence assumption. 

While the attention on the full sequence in AED is a natural solution to machine translation which has the word order switching between source and target languages, it may not be ideal to ASR because the speech signal and output label sequence are monotonic. In order to have better alignment between the speech signal and label sequence, the AED model is optimized together with a CTC model in a multi-task learning framework by sharing the encoder \cite{kim2017joint}.  Such a training strategy greatly improves the convergence of the attention-based model and mitigates the alignment issue. It becomes the standard training recipe for most AED models \cite{watanabe2017hybrid, hori2017advances, toshniwal2017multitask, ueno2018acoustic, delcroix2018auxiliary, liu2019adversarial, kim2019attention, karita2019improving}.  In \cite{hori2017joint}, a further improvement was proposed by combining the scores from both the AED model and the CTC model during decoding.

\begin{figure}
     \centering
     \begin{subfigure}[b]{0.22\textwidth}
         \centering
         \includegraphics[width=\textwidth]{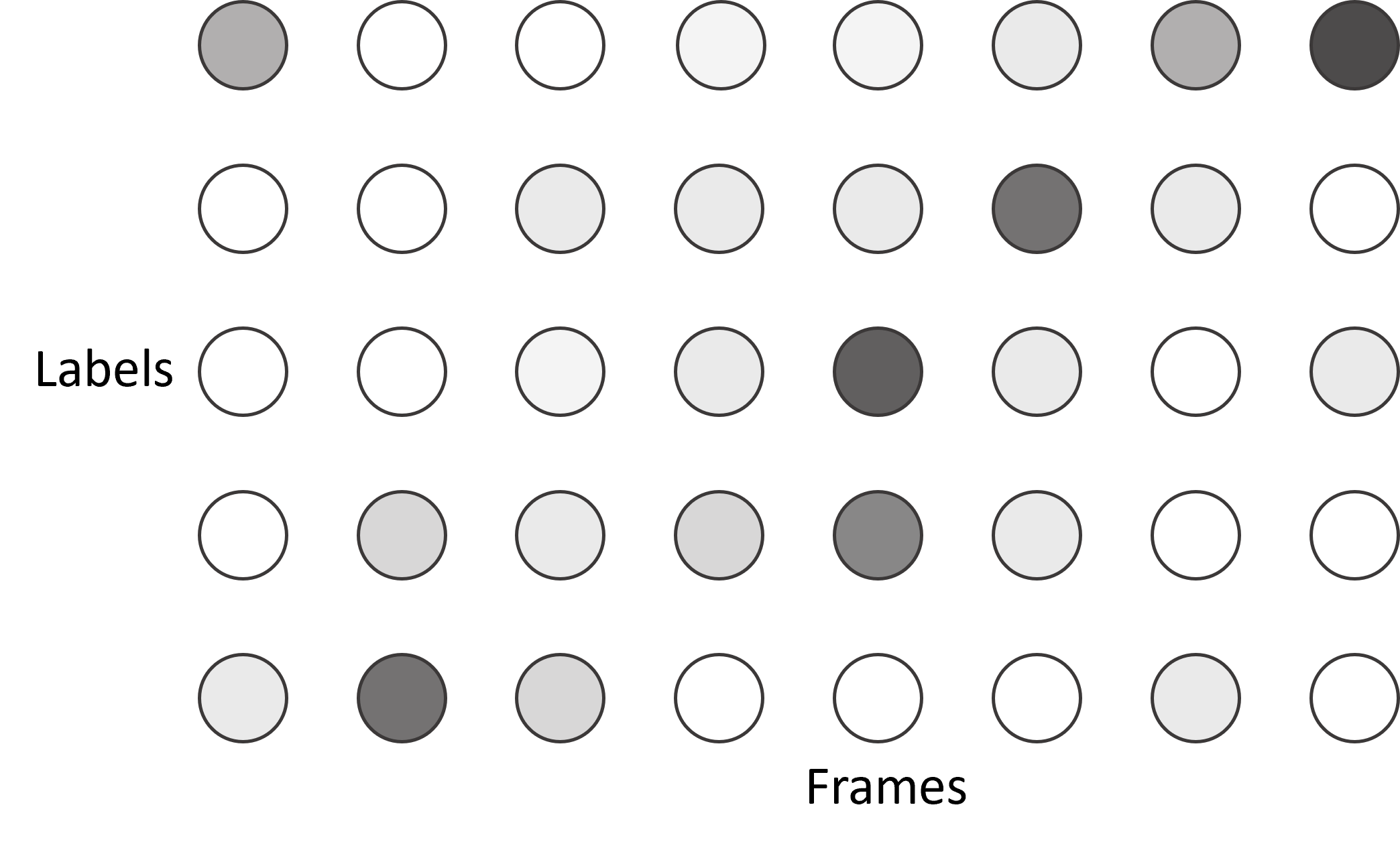}
         \caption{full attention}
         \label{fig:full_att}
     \end{subfigure}
     \hfill
     \begin{subfigure}[b]{0.22\textwidth}
         \centering
         \includegraphics[width=\textwidth]{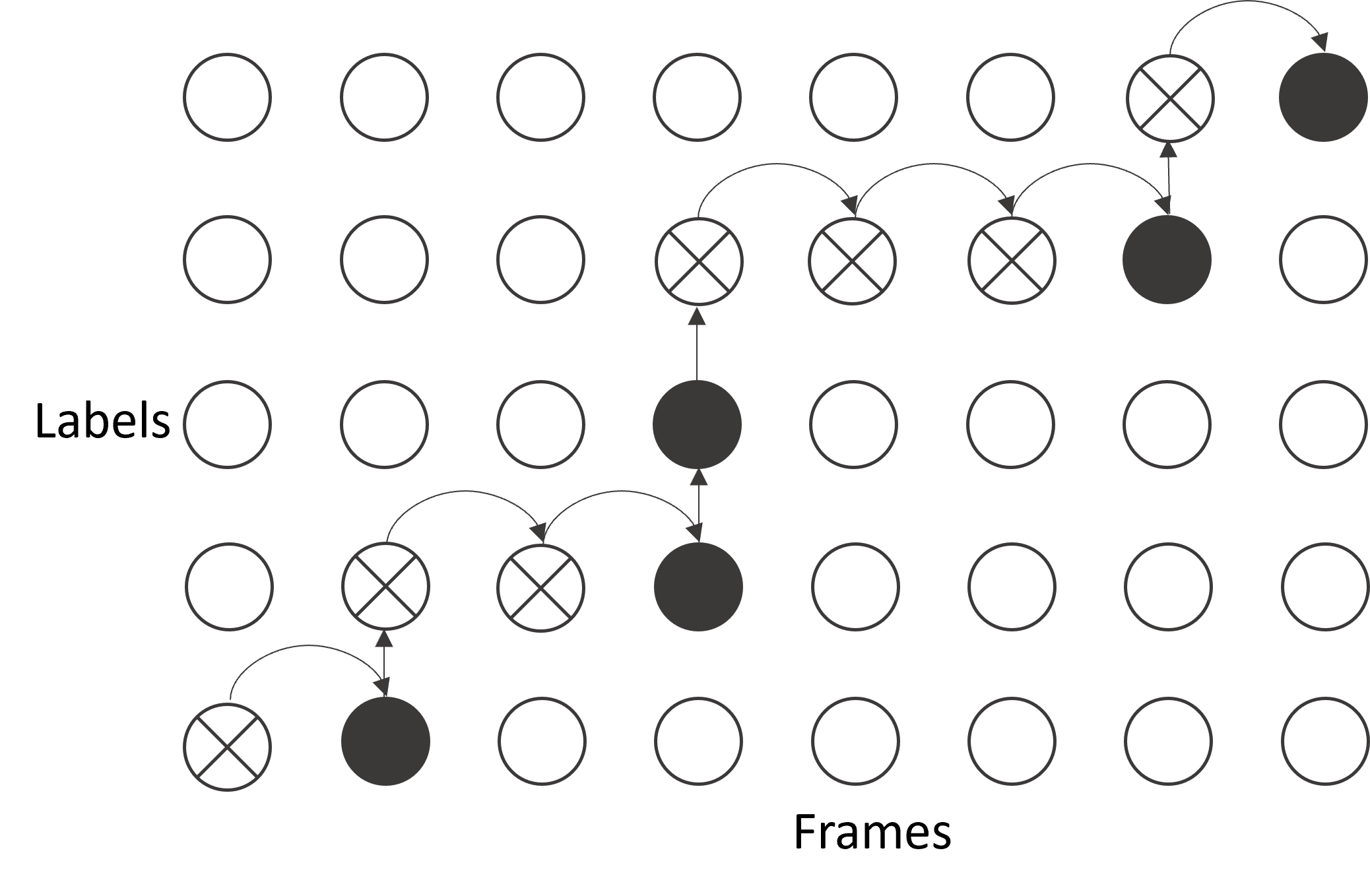}
         \caption{monotonic attention}
         \label{fig:mono}
     \end{subfigure}
     \hfill
     \begin{subfigure}[b]{0.22\textwidth}
         \centering
         \includegraphics[width=\textwidth]{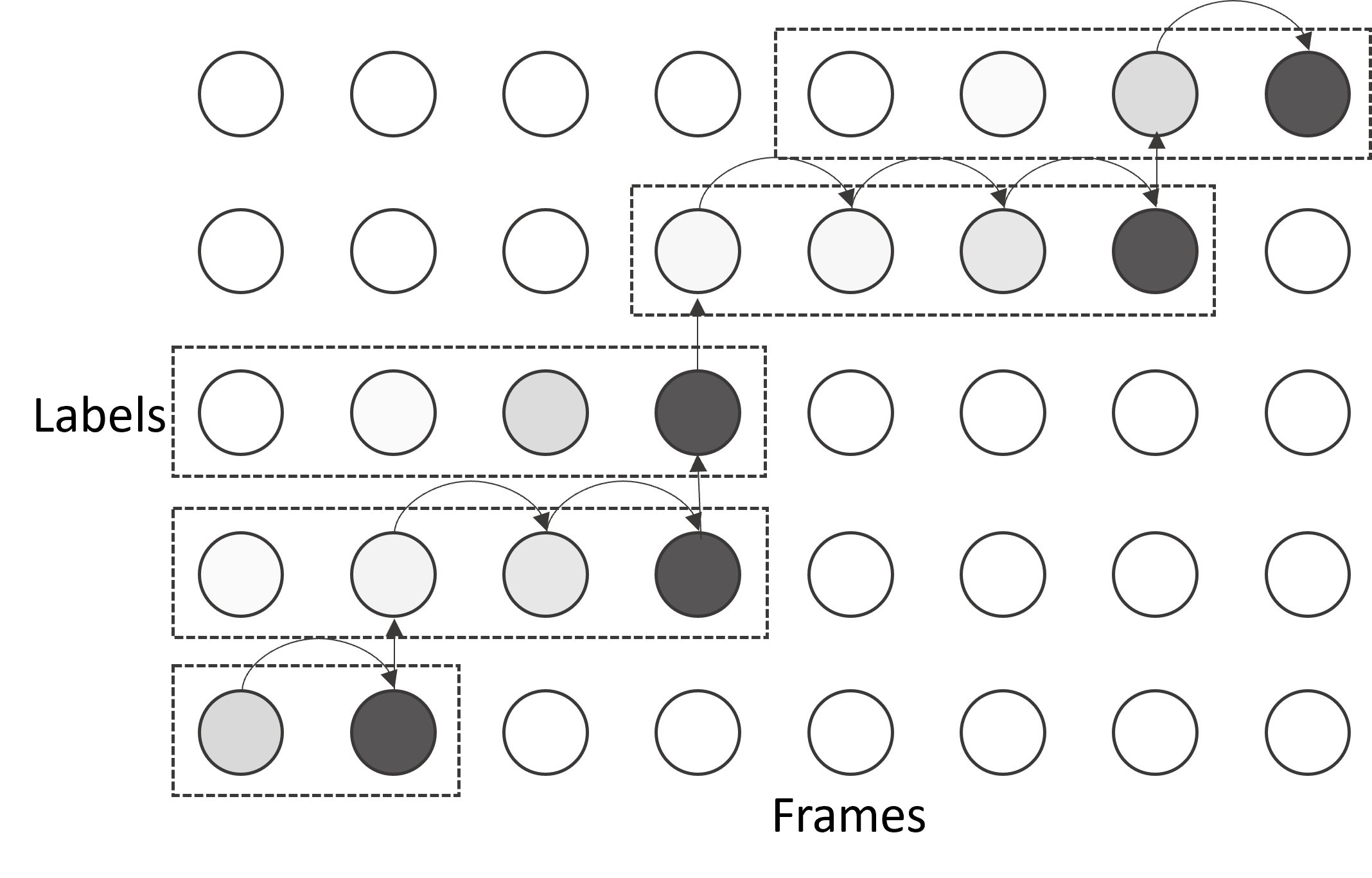}
         \caption{MoChA}
         \label{fig:mocha}
     \end{subfigure}
     \hfill
     \begin{subfigure}[b]{0.22\textwidth}
         \centering
         \includegraphics[width=\textwidth]{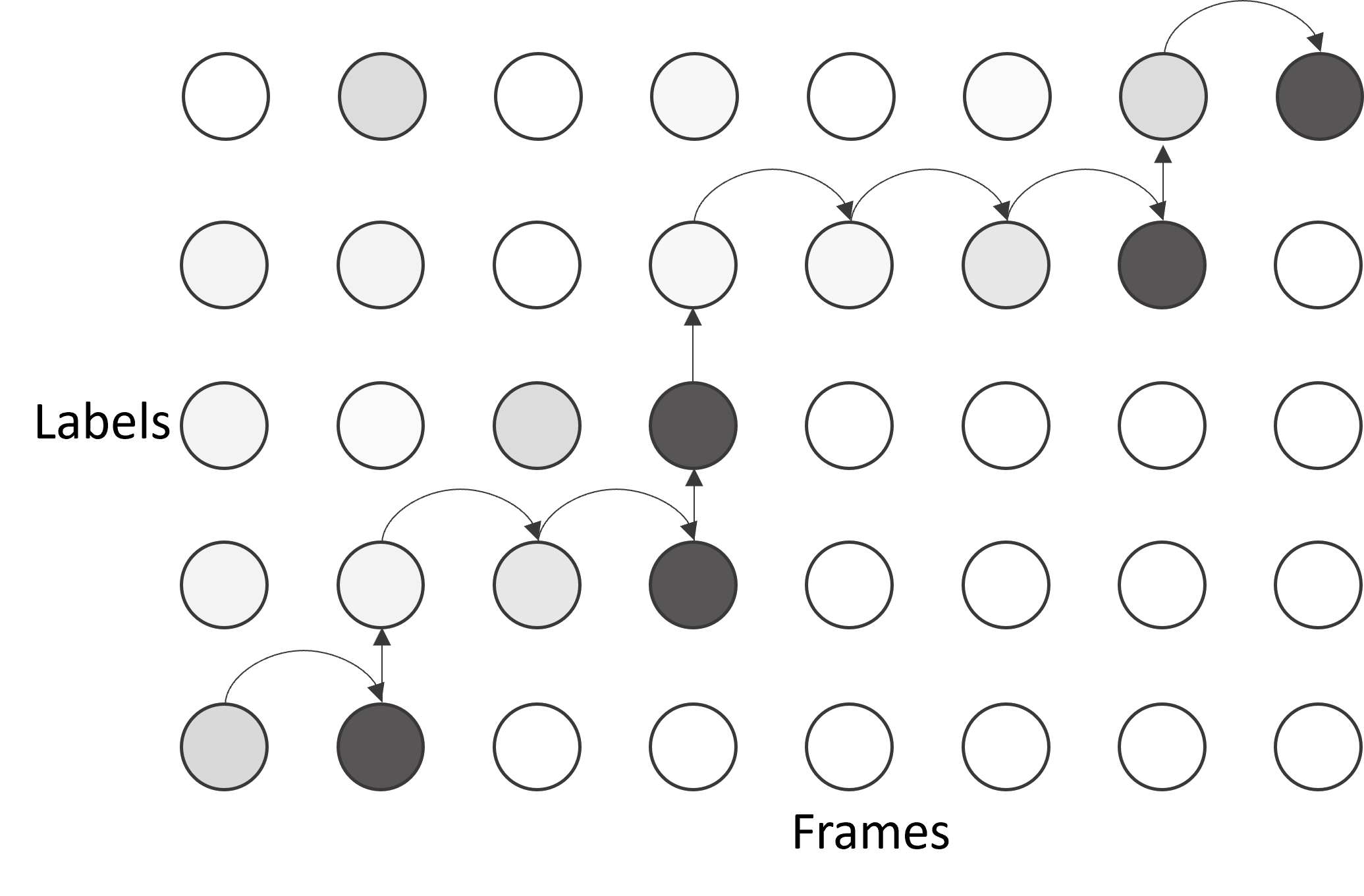}
         \caption{MILK}
         \label{fig:milk}
     \end{subfigure}     
        \caption{Attention methods for AED. The time step goes from left to right, and the label sequence goes from bottom to top. Every circle represents an attention weight. The darker the circle is, the larger the attention weight is. }
        \label{fig:attention_AED}
\end{figure}

Most commercial setups need ASR systems to be streaming with low latency, which means ASR systems should produce the recognition results at the same time as the user is speaking. In vanilla AED models, the attention is applied to the whole utterance in order to achieve good performance. The latency can be significant because such a setup needs to obtain the full utterance before decoding, and it is impractical for streaming ASR scenarios where the speech signal comes in a continuous mode without segmentation. There are lots of attempts to build streaming  AED models. The basic idea of these methods is applying attention on the chunks of the input speech. The difference between these attempts is the way the chunks are determined and used for attention. 
Figure \ref{fig:full_att} shows the full attention of AED, which spreads attention weights across the whole utterance. 
One major school of streaming AED models is to use monotonic attention. In \cite{raffel2017online}, a monotonic attention mechanism was proposed to use integrated decision for triggering ASR. Attention is only on the time step corresponding to the trigger point as shown in Figure \ref{fig:mono}. 
The method was improved in \cite{chiu2018monotonic}, where a monotonic chunkwise attention (MoChA) method was proposed to stream the attention by splitting the encoder outputs into small fixed-size chunks so that the soft attention is only applied to those small chunks, as shown in Figure \ref{fig:mocha}. Adaptive-size instead of fixed-size chunks were used in \cite{fan2019online}. 
Recently, monotonic infinite lookback (MILK) attention \cite{arivazhagan2019monotonic} was proposed to attend  the entire sequence preceding the trigger point, as shown in Figure \ref{fig:milk}.
Another popular streaming AED method is triggered attention \cite{moritz2019triggered} which uses CTC segments to decide the chunks and is applied to lots of AED models \cite{moritz2019streaming, moritz2020streaming,  hori2021advanced}. 

While all these methods can achieve the goal of streaming ASR to some extent, they usually do not enforce low-latency which is another important factor for commercial ASR systems. This challenge was addressed in \cite{inaguma2020minimum}, which proposed to train low-latency streaming AED models by leveraging the external hard alignment. In \cite{wang2020low}, a scout network was used to predict word boundary which is then used by the ASR network to predict the next subword by utilizing the information from all speech frames before it for low latency streaming. 

All streaming AED models need some complicated strategies to decide the trigger point for streaming. A comparison of  streaming ASR methods was conducted in \cite{kim2021comparison} which shows RNN Transducer has advantages over MoChA in terms of latency, inference time, and training stability.  AED models also cannot perform well on long utterances \cite{chiu2019comparison, narayanan2019recognizing}. Therefore, although there are still many activities \cite{li2019end, kim2019attention, dong2020cif, miao2020transformer, tian2020synchronous, moritz2020streaming, inaguma2020enhancing, li2021transformer, tsunoo2021streaming, sterpu2021learning,  kashiwagi2021gaussian} in the area of streaming AED models, the industry tends to choose RNN Transducer introduced next as the dominating streaming E2E model while AED has its position in some non-streaming scenarios. 

\subsection{RNN Transducer} \label{ssec:RNN-T} 

RNN Transducer (RNN-T) \cite{Graves-RNNSeqTransduction} provides a natural way for streaming ASR because its output conditions on the previous output tokens and the speech sequence until the current time step (i.e., frame) as $P(y_u | {\bf{x}}_{1:t} , {\bf{y}}_{1:u-1})$. In this way, it also removes the conditional independence assumption of CTC.  With the natural streaming capability, RNN-T becomes the most popular E2E model in the industry \cite{prabhavalkar2017comparison, battenberg2017exploring, he2019streaming, Li2019improving, sainath2020streaming, Li2020Developing, zhang2021benchmarking, saon2021advancing, punjabi2021joint}. 

Illustrated in Figure \ref{fig:RNNT}, RNN-T contains an encoder network, a prediction network, and a joint network.  
The encoder network is the same as that in CTC and AED, generating a high-level feature representation ${\bf{h}}_t^{enc}$. The prediction network produces a high-level representation ${\bf{h}}_u^{pre}$ based on RNN-T's previous output label $y_{u-1}$. The joint network is a feed-forward network that combines ${\bf{h}}_t^{enc}$ and  ${\bf{h}}_u^{pre}$ as 
\begin{equation}
{\bf{z}}_{t,u}  = \psi({\bf{Q}} {\bf{h}}_t^{enc} + {\bf{V}} {\bf{h}}_u^{pre} + {\bf{b}}_z), \label{eq:z}
\end{equation}
where ${\bf{Q}}$ and ${\bf{V}}$ are weight matrices, ${\bf{b}}_z$ is a bias vector, and $\psi$ is a non-linear function (e.g., RELU or Tanh).
${\bf{z}}_{t,u}$ is connected to the output layer with a linear transform
\begin{align}
{\bf{h}}_{t,u}={\bf{W}}_y {\bf{z}}_{t,u} +{\bf{b}}_y,\label{eq:rnnt_h}
\end{align}
where ${\bf{W}}_y$ and ${\bf{b}}_y$ denote a weight matrix and a bias vector, respectively. The probability of each output token $k$ is
\begin{align}
P(y_u=k|{\bf{x}}_{1:t},{\bf{y}}_{1:u-1})=softmax (h_{t,u}^k).
\label{eq:rnnt_soft}
\end{align}

The loss function of RNN-T is also $- ln P( {\bf{y}|\bf{x}} )$,
with 
\begin{equation} \label{eq:Prob_RNNT}
P( {\bf{y}|\bf{x}} ) = \sum_{{\bf{a}} \in A^{-1}(\bf{y})} P( \bf{a} | \bf{x} )
\end{equation}
as the sum of all possible alignment paths that are mapped to the label sequence $\bf{y}$. The mapping from the alignment path $\bf{a}$ to the label sequence $\bf{y}$ is defined as  $A (\bf{a}) = \bf{y}$. \footnote{Note this mapping is different from the CTC mapping $B$ in Eq. \eqref{eq:Prob_CTC}.}

\begin{figure}[t]
  \centering
  \includegraphics[width=0.9\linewidth]{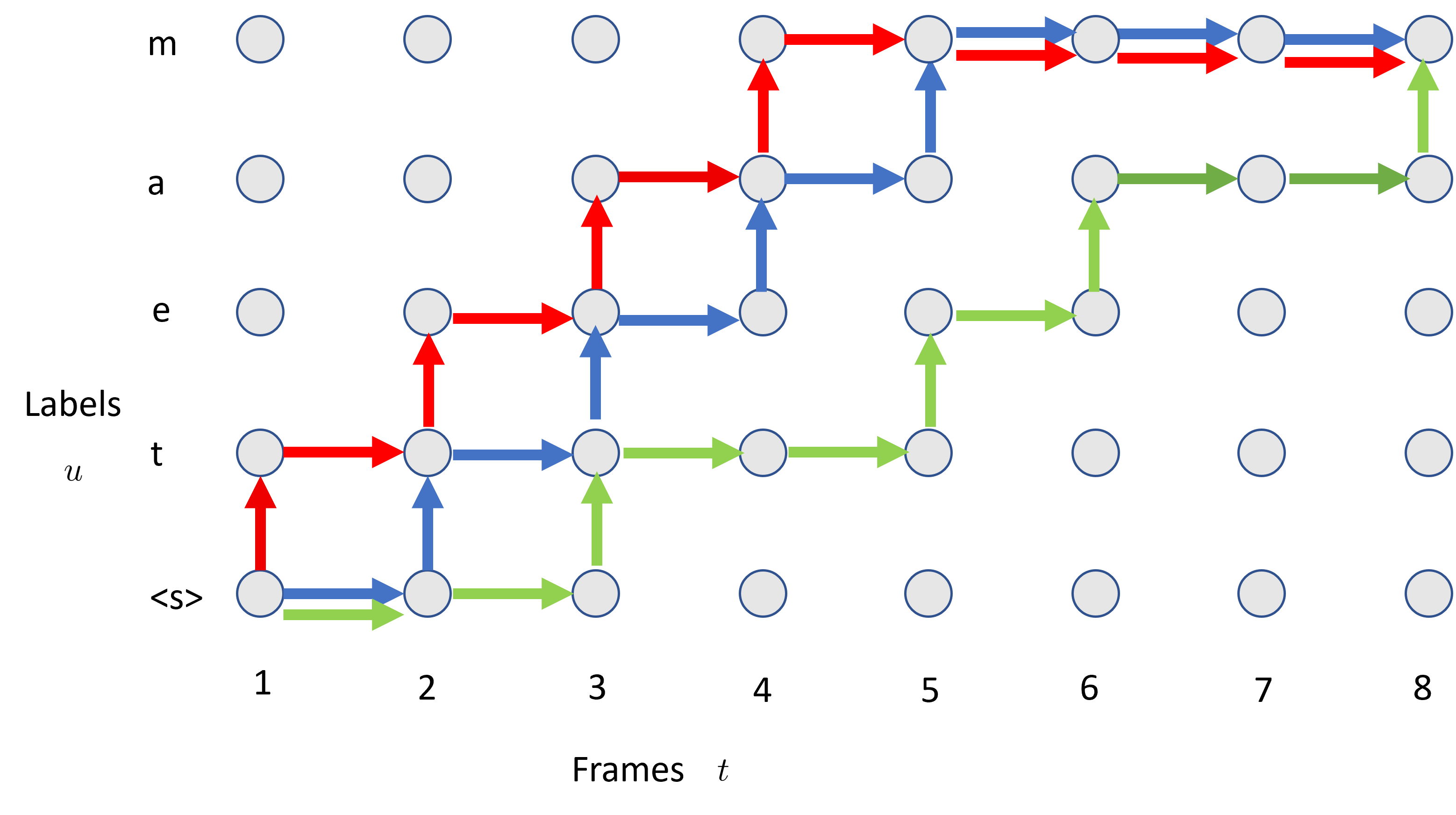}
  \caption{Example alignment paths of RNN-T. }
  \label{fig:RNNT_path}
\end{figure}

In Figure \ref{fig:RNNT_path}, three example alignment paths are plotted for speech sequence ${\bf{x}}=({\bf{x_1}}, {\bf{x_2}},......,{\bf{x_8}})$ and label sequence ${\bf{y}}=(\langle s \rangle, t, e, a, m)$, where $\langle s \rangle$ is a token for sentence start. All valid alignment paths go from the bottom left corner to the top right corner of the $T\text{x}U$ grid, hence the length of  each alignment path is $T+U$. In an alignment path, the horizontal arrow advances one time step with a blank label by retaining the prediction network state while the vertical arrow emits a non-blank output label. 

The posteriors of the alignment grid composed by the encoder and prediction networks need to be calculated at each grid point. This a three-dimensional tensor that requires much more memory than what is needed in the training of other E2E models such as CTC and AED. Eq. \eqref{eq:Prob_RNNT} is calculated based on the forward-backward algorithm described in \cite{Graves-RNNSeqTransduction}. In \cite{bagby2018efficient}, to improve training efficiency, the forward and backward probabilities can be vectorized with a loop skewing transformation, and the recursions can be computed in a single loop instead of two nested loops. Function merging was proposed in  \cite{Li2019improving} to significantly reduce the training memory cost so that larger minibatches can be used to improve the training efficiency. 

It is worth noting that ASR latency is a very important metric that affects user experience. If the latency is large, users will feel the ASR system is not responding. Therefore, practical systems need to have a small latency in order to give users a good experience \cite{sainath2021efficient}. Shangguan et al. recently gave a very good study of the factors affecting user perceived latency in \cite{shangguan2021dissecting}.
In \cite{chang2019joint}, RNN-T is designed to generate end of sentence (EOS) token together with the ASR transcription. The latency of EOS detection is improved in \cite{li2020towards} with early and late penalties. While RNN-T is now the most natural E2E model for streaming, the vanilla RNN-T still has latency challenges because RNN-T tends to delay its label prediction until it is very confident by visiting more future frames of the current label. The green path in Figure \ref{fig:RNNT_path} is an example alignment of such a delayed decision. In order to ensure low latency for RNN-T, constrained alignment \cite{sainath2020emitting, mahadeokar2021alignment} was proposed to restrict the training alignment within a delay threshold of the ground truth time alignment and forbid other alignment paths. 
Strict alignment between the input speech sequence and output label sequence is enforced in \cite{tripathi2019monotonic} to generate better alignment for streaming RNN-T. In addition to the benefit of latency, another advantage of all these alignment restricted RNN-T methods is GPU-memory saving and training speed-up because less alignment paths are used during training.     FastEmit \cite{yu2021fastemit} was proposed to reduce latency without the need of the ground truth alignment. It encourages the emission of vocabulary tokens and discourages the blank transition at all positions in the time-label grid. Therefore, it pushes all  alignment paths to the left direction of the time-label grid. However, this operation is a little aggressive at reducing latency, which can result in recognition accuracy loss. In \cite{kim2021reducing}, a self-alignment method was proposed to reduce latency in a mild way. It uses the self alignment of the current model to find the lower latency alignment direction. In Figure \ref{fig:RNNT_path}, the blue path indicates a self-alignment path and the red path is one frame left to the self-alignment path. During training, the method encourages the left-alignment path, pushing the model's alignment to the left direction. 
The proposed method was reported to have better accuracy and latency tradeoff than the constrained-alignment method and FastEmit. 

In addition to these three popular E2E models, there are also other E2E models such as neural segmental model \cite{tang2017end} and recurrent neural aligner \cite{sak2017recurrent}, to name a few. 


\section{Encoder}
In all E2E ASR models, the most important component is the encoder which converts speech input sequences into high-level feature representations. 

\subsection{LSTM}
In early E2E ASR works, LSTM is the most popular model unit. The encoder can be either a  multi-layer unidirectional LSTM-RNN as  Eq. \eqref{eq:uni_enc} 
\begin{equation}
{\bf{h}}_t^{l} = \text{LSTM} ({\bf{x}}_t^l, {\bf{h}}_{t-1}^l)	\label{eq:uni_enc}
\end{equation}
or a multi-layer  bidirectional LSTM (BLSTM)-RNN as Eq. \eqref{eq:bi_enc}
\begin{equation}
{\bf{h}}_t^{l} = [\text{LSTM} ({\bf{x}}_t^l, {\bf{h}}_{t-1}^{l}), \text{LSTM} ({\bf{x}}_t^l, {\bf{h}}_{t+1}^{l})],	\label{eq:bi_enc}
\end{equation}
where $\text{LSTM}()$ denotes the standard LSTM unit with an optional projection layer \cite{Sak2014long}. Here, ${\bf{h}}_{t}^l$ is the hidden output of the $l$-th ($l=1...L$) layer at time $t$ and ${\bf{x}}_{t}^l$ is the input vector for the $l$-th layer with
 \begin{equation}
    {\bf{x}}_{t}^l = 
\begin{cases}
    {\bf{h}}_t^{l-1},& \text{if } l > 1 \label{eq:lstmx} \\
    {\bf{x}}_t,              & \text{if } l = 1
\end{cases},
\end{equation}
where ${\bf{x}}_t$ is the speech input at time step $t$. The last layer output, ${\bf{h}}_t^{L}$, is used as the encoder output. 

Because of the streaming request in most commercial ASR systems, unidirectional LSTM is used more widely. When the encoder is an LSTM-RNN, CTC and RNN-T work in the streaming mode by default while AED still needs streaming attention strategies in order to work in the streaming mode. In contrast, when the encoder is a BLSTM-RNN, all E2E models are non-streaming models. 

There is a clear accuracy gap between the LSTM encoder and the BLSTM encoder because the latter uses the whole utterance information to generate the encoder output. In order to reduce the gap, it is a natural idea to use future context frames to generate more informative encoder output with the methods such as latency controlled BLSTM (LC-BLSTM) \cite{zhang2016highway} or contextual LSTM (cLSTM) \cite{li2019improvinglayer, Li2020Developing}. LC-BLSTM works in an overlapped multi-frame chunk mode. BLSTM is still used within a chunk, while the state of the forward direction is carried across chunks. In contrast, cLSTM works in a frame-by-frame uni-directional way by taking the lower layer's output from future frames as the input of the current frame.

\subsection{Transformer}
\begin{figure}
     \centering
     \begin{subfigure}[b]{0.18\textwidth}
         \centering
         \includegraphics[width=\textwidth]{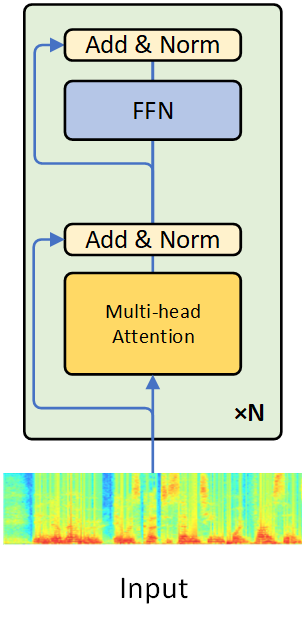}
         \caption{Transformer}
         \label{fig:Transformer}
     \end{subfigure}
     \hfill
     \begin{subfigure}[b]{0.28\textwidth}
         \centering
         \includegraphics[width=\textwidth]{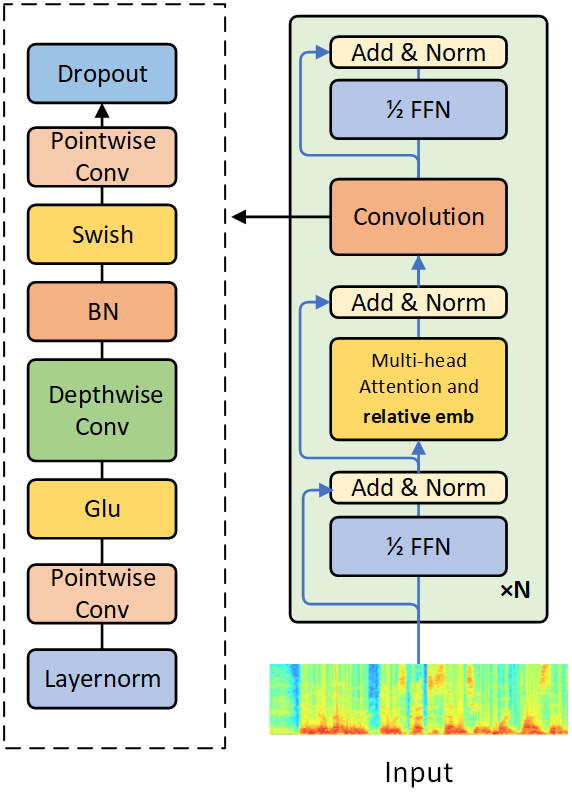}
         \caption{Conformer}
         \label{fig:Conformer}
     \end{subfigure}
        \caption{The structures of Transformer and Conformer}
        \label{fig:Trans_Conformer}
\end{figure}

Although LSTM can capture short-term dependencies, Transformer is much better at capturing long-term dependencies because its attention mechanism sees all context directly. Transformer was shown to outperform LSTM \cite{zeyer2019comparison, karita2019comparative, li2020comparison} and is now replacing LSTM in all E2E models \cite{dong2018speech, karita2019comparative, zeyer2019comparison, zhang2020transformer, li2020comparison, higuchi2020mask}.  
The encoder of Transformer-based E2E models is composed of a stack of Transformer blocks, where each block has a multi-head self-attention layer and a feed-forward network (FFN), as shown in Figure \ref{fig:Transformer}. Residual connections \cite{he2016deep} and layer normalization  \cite{ba2016layer} are used to connect different layers and blocks.  The input of a Transformer block can be linearly transformed to the query $\bf{q}$, key $\bf{k}$, and value $\bf{v}$ vectors with matrices $\bf{W}_q$, $\bf{W}_k$, and $\bf{W}_v$, respectively.
Self-attention is used to compute the attention distribution over the input speech sequence with the dot-product similarity function as
\begin{eqnarray} 
    \alpha_{t, \tau} &=& \frac{\exp(\beta ({\bf{W}}_q {\bf{x}}_t)^T({\bf{W}}_k{\bf{x}}_\tau))}{\sum_{\tau'}\exp(\beta ({\bf{W}}_q{\bf{x}}_t)^T ({\bf{W}}_k{\bf{x}}_{\tau'}))}  \nonumber \\ 
    &=& softmax (\beta {\bf{q}}_t^T {\bf{k}}_\tau),
    \label{eqn:self_att}
\end{eqnarray}
where $\beta=\frac{1}{\sqrt{d}}$ is a scaling factor and $d$ is the dimension of the feature vector for each head. Then, the attention weights are used to combine the value vectors to generate the layer output at the current time step
\begin{equation} 
        {\bf{z}}_t = \sum_{\tau} \alpha_{t\tau} {\bf{W}}_v {\bf{x}}_\tau = \sum_{\tau} \alpha_{t\tau} {\bf{v}}_\tau. 
\end{equation}
Multi-head self-attention (MHSA) is used to further improve the model capacity by applying multiple parallel self-attentions on the input sequence and the outputs of each attention module are then concatenated. 
By replacing the LSTM modules with Transformer in the encoder, the RNN-T model becomes Transformer Transducer \cite{yeh2019transformer, zhang2020transformer, chen2021developing, dalmia2021transformer} which has better accuracy than RNN-T due to the modeling power of Transformer. 

While Transformer is good at capturing global context, it is less effective in extracting local patterns. To further improve the modeling capability, convolutional neural network (CNN) \cite{waibel1989phoneme, lecun1989backpropagation} which works on local information is combined with Transformer as Conformer \cite{gulati2020conformer} which is plotted in Figure \ref{fig:Conformer}. The conformer block contains two half-step FFNs sandwiching the MHSA and convolution modules. Although having higher model complexity than  Transformer, Conformer has been reported to outperform Transformer in various tasks \cite{gulati2020conformer, guo2021recent}. 

Again, the self-attention on the full speech sequence cannot be used for streaming ASR. In \cite{chen2021developing}, a masking strategy was proposed to flexibly control how attention is performed. A binary attention mask $m_{t,\tau}$ is introduced to the self-attention in Eq. \eqref{eqn:self_att} as 
\begin{eqnarray} 
    \alpha_{t, \tau} &=& \frac{m_{t,\tau} \exp(\beta ({\bf{W}}_q {\bf{x}}_t)^T({\bf{W}}_k{\bf{x}}_\tau))}{\sum_{\tau'} m_{t,\tau'} \exp(\beta ({\bf{W}}_q{\bf{x}}_t)^T ({\bf{W}}_k{\bf{x}}_{\tau'}))}  \nonumber \\ 
    &=& softmax (\beta {\bf{q}}_t^T {\bf{k}}_\tau, m_{t,\tau}).
    \label{eqn:self_att_mask}
\end{eqnarray}

\begin{figure}
     \centering
     \begin{subfigure}[b]{0.5\textwidth}
         \centering
         \includegraphics[width=\textwidth]{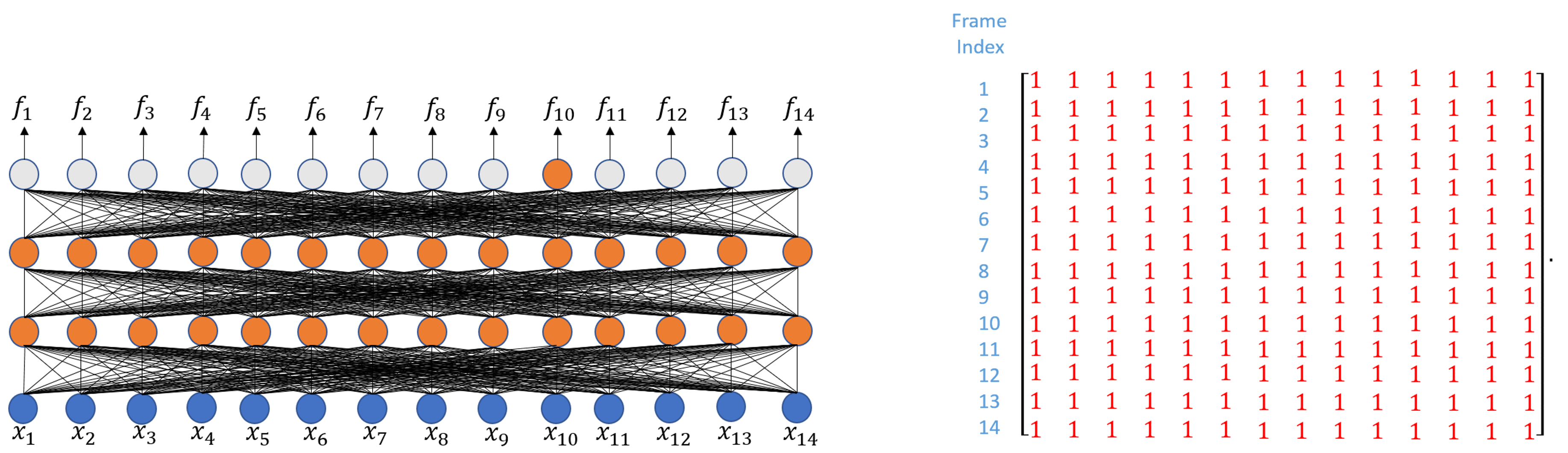}
         \caption{full attention}
         \label{fig:TT_full}
     \end{subfigure}
     \hfill
     \begin{subfigure}[b]{0.5\textwidth}
         \centering
         \includegraphics[width=\textwidth]{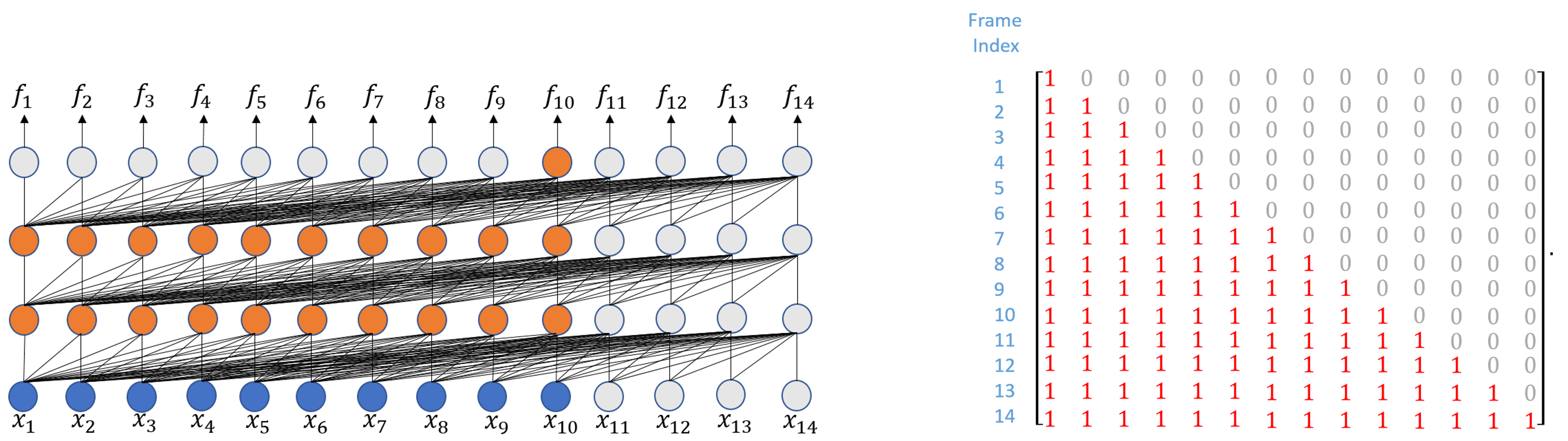}
         \caption{attention on whole history}
         \label{fig:TT_AllHistory}
     \end{subfigure}
     \hfill
     \begin{subfigure}[b]{0.5\textwidth}
         \centering
         \includegraphics[width=\textwidth]{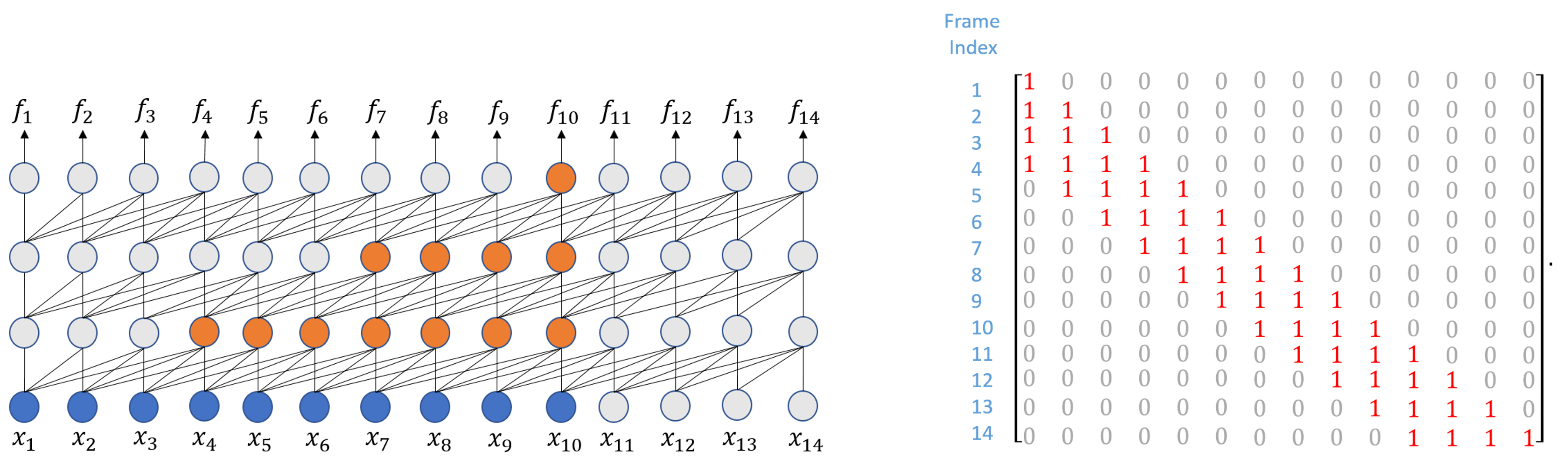}
         \caption{attention with context expansion for history frames}
         \label{fig:TT_LayerHistory}
     \end{subfigure}
     \hfill
     \begin{subfigure}[b]{0.5\textwidth}
         \centering
         \includegraphics[width=\textwidth]{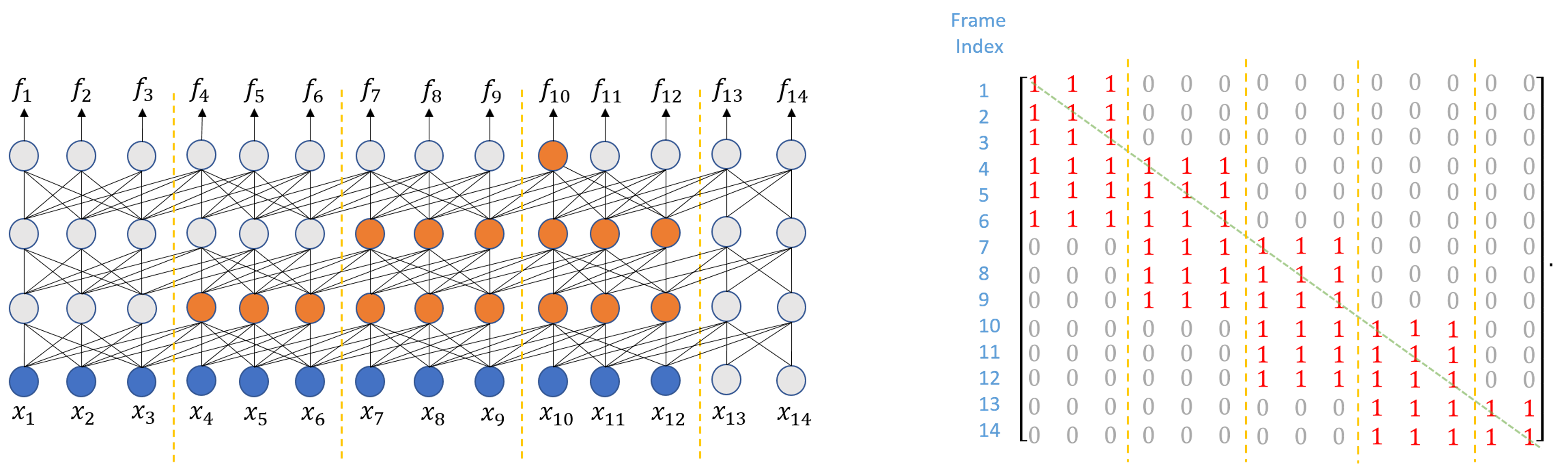}
         \caption{attention with context expansion for history frames and a future chunk}
         \label{fig:TT_MS}
     \end{subfigure}     
        \caption{Attention for different computational cost and latency configuration when predicting the output for ${\bf{x}}_{10}$. In every sub figure, the left side is the reception field, and the right side is the attention mask matrix.}
        \label{fig:TT_attention}
\end{figure}

Figure \ref{fig:TT_attention} shows several examples of how the binary mask is used for different computational costs and latency configurations when predicting the output for ${\bf{x}}_{10}$.  The full attention case is plotted in Figure \ref{fig:TT_full}, where the reception field is the full sequence, and every element of the mask matrix is $1$. For the strict streaming setup without any latency, the attention should not be performed on any future frame, as shown in Figure \ref{fig:TT_AllHistory}. The major drawback of this attention mechanism is that the memory and runtime cost increases linearly as the utterance history grows. The context operation \cite{zhang2020transformer} was proposed to address the issue by working on a limited history at every layer. Shown in \ref{fig:TT_LayerHistory}, because of context expansion, the model still can access a very long history. If the attention operates on $F$ history frames at every layer and the model has $L$ Transformer layers, then the model can access $F\text{x}L$ history frames. In order to improve modeling accuracy, it is better to also include some future frames into the prediction. In \cite{zhang2020transformer}, the same context expansion strategy is applied to not only the history but also the future frames. However, because of context expansion, it needs a large amount of future frames when there are many Transformer layers, introducing significant latency. A better future access strategy was proposed in  \cite{chen2021developing, shi2021emformer} where the input speech sequence is segmented into fixed-size chunks. The frames in the same chunk can see each other and each frame can see fixed numbers of the left chunks so that the left reception field propagates. Plotted in Figure \ref{fig:TT_MS}, this strategy helps to build high accuracy models with a very small latency. In addition to these methods, there are lots of variations \cite{tsunoo2019transformer, moritz2020streaming, wang2020low, tian2020synchronous, zhang2020streaming, huang2020conv} of localizing full self-attention in order to have a streaming Transformer encoder.

Another popular method is Gaussian soft-mask \cite{sperber2018self} which changes the softmax calculation in self-attention in Eq. \eqref{eqn:self_att} as 
\begin{equation}
\alpha_{t, \tau} = softmax (\beta {\bf{q}}_t^T {\bf{k}}_\tau+ M_{t, \tau}),    
\end{equation}
where $M_{t, \tau}$ is a mask. It can be either a hard mask which sets all weights outside the band $b$ to 0 as 
\begin{equation}
M_{t, \tau}=
    \begin{cases}
      0, & \text{if}\ |t-\tau|<\frac{b}{2} \\
      -\infty, & \text{otherwise}
    \end{cases}
\end{equation}
or a soft mask as 
\begin{equation}
M_{t, \tau}=\frac{-(t-\tau)^2}{2\sigma}, 
\end{equation}
where $\sigma$ is a trainable parameter. While both masks can enforce the locality which is good for the speech signal, the soft-mask method cannot reduce computational cost and latency because it sill attends the full sequence, although the weights of far-away time steps are small.

\section{Other Training Criterion}
In addition to the standard training loss  $- ln P( {\bf{y}}|{\bf{x}} )$ for E2E models, there are other losses used in the popular  teacher-student learning and minimum word error rate training.

\label{sec:training}

\subsection{Teacher-Student Learning}
The concept of teacher-student (T/S) learning was originally introduced in \cite{bucilua2006model} but became popular from the deep learning era. The most popular T/S learning strategy is to minimize the KL divergence between the output distributions of the teacher and student networks, first proposed in 2014 by Li et al. \cite{li2014learning}. Later, Hinton et al. \cite{hinton2015distilling} branded it with a new name, knowledge distillation, by introducing a temperature parameter (like chemical distillation) to scale the posteriors. 

In the context of E2E modeling, the token-level loss function of T/S learning is
\begin{equation}    \label{eq:ts}
    -\sum_u\sum_k P_T(k \mid {\bf{y}}_{1:u-1}, {\bf{x}}) \log P_S(k \mid {\bf{y}}_{1:u-1}, {\bf{x}}),
\end{equation}
where $P_T$ and $P_S$ denote the posterior distributions of the teacher and student networks, respectively. The most popular usage of T/S learning in E2E modeling is to learn a small student E2E model from a large teacher E2E model \cite{li2018developing, pang2018compression, kim2019knowledge, panchapagesan2021efficient}. 

Another usage is to learn a streaming E2E model from a non-streaming E2E model \cite{kim2018improved, kurata2019guiding, moriya2020distilling, kurata2020knowledge, kojima21knowledge}. It is a little tricky here because the streaming E2E models usually generate delayed decisions. For example, the streaming CTC and non-streaming CTC have different output spike patterns. Therefore, delicate work has been done in order to make the non-streaming E2E models generate friendly output spikes that can be learned by the streaming E2E models \cite{kurata2019guiding, kurata2020knowledge}.  Another way to circumvent the spike mismatch is to train the streaming E2E model by using the ASR hypotheses generated from decoding the unlabeled data with a non-streaming E2E model  \cite{doutre2021improving}.

T/S learning can also be used to adapt an E2E model from a source environment to a target environment if the target-domain acoustic data $\hat{\bf{x}}$  can be paired with the source-domain acoustic data $\bf{x}$ either by recording at the same time or from simulation \cite{meng2019domain, zhang2019knowledge}. The loss function in Eq. \eqref{eq:ts} is re-written as
\begin{equation}    \label{eq:ts_domain}
    -\sum_u\sum_k P_T(k \mid {\bf{y}}_{1:u-1}, {\bf{x}}) \log P_S(k \mid {\bf{y}}_{1:u-1}, \hat{\bf{x}}).
\end{equation}

While standard T/S learning follows a two-step process in which we first train a teacher model and then use the T/S criterion to train a student model,  there are recent studies \cite{wu2021dynamic, yu2021dual, swaminathan2021codert, nagaraja2021collaborative} that train the teacher and student models at the same time. Such a co-learning strategy not only simplifies the training process but also boosts the accuracy of student models. 

\subsection{Minimum Word Error Rate Training}
\label{ssec:MWER}
The minimum word error rate (MWER) training criterion tries to mitigate the discrepancy between the training criterion and the evaluation metric of a speech recognition model. Instead of minimizing the negative log loss of the probability $P( {\bf{y}|\bf{x}} )$, MWER minimizes the expected word errors \cite{shannon2017optimizing} as 
\begin{align} 
\mathcal{L}_{\text{MWER}} = \sum_{{\bf{h}}_i} \hat{P}({\bf{h}}_i \mid {\bf{x}}) R({\bf{h}}_i, {\bf{h}}^r),
\end{align}
where $\hat{P}({\bf{h}}_i \mid {\bf{x}})$ denotes the posterior probability of an hypothesis ${\bf{h}}_i$, $R(\cdot)$ denotes the risk function which measures the edit-distance between the hypothesis ${\bf{h}}_i$ and the reference transcription ${\bf{h}}^r$ at word-level. While the exact posterior probability is computationally intractable, in practice, an N-best list of hypotheses from beam search decoding are used to compute the empirical posterior probability \cite{prabhavalkar2018minimum} as
\begin{align}
\label{eq:post}
\hat{P}({\bf{h}}_i \mid {\bf{x}}) = \frac{P({\bf{h}}_i \mid {\bf{x}})}{\sum_{{\bf{h}}_i} P({\bf{h}}_i | {\bf{x}})},
\end{align}
where $P({\bf{h}}_i \mid {\bf{x}})$ is the probability of an hypothesis ${\bf{h}}_i$, computed in Eq. \eqref{eq:Prob_attention} for AED models or Eq. \eqref{eq:Prob_RNNT} for RNN-T models. 

MWER has been shown to improve the accuracy of AED models \cite{prabhavalkar2018minimum, cui2018improving}, RNN-T models \cite{weng2020minimum, guo2020efficient}, and hybrid autoregressive transducer (HAT) models \cite{lu2021on}. However, the gain is not as significant as what has been reported in hybrid models. One possible reason is that the MWER training of hybrid models follows the cross-entropy training which is frame-based. In contrast, E2E models have already been optimized with sequence-level training, therefore further sequence-level MWER training gives less gain.

\section{Multilingual modeling}
Building a multilingual ASR system with traditional hybrid models is difficult because every language usually has its own language-specific phonemes and word inventories. Most hybrid multilingual works focus on building an acoustic model with shared hidden layers \cite{huang2013cross, heigold2013multilingual, ghoshal2013multilingual}, while every language has its own lexicon model and language model. In contrast, it is very easy to build a multilingual E2E ASR system by simply taking a union of the token (e.g., character,  subword, or even byte) sets of all languages as the output token set and then training an E2E model with all the data \cite{watanabe2017language, kim2018towards, toshniwal2018multilingual, cho2018multilingual, li2019bytes}. Such an E2E model is a universal ASR model which is capable of recognizing the speech from any language as long as that language has been used during training.  However, pooling all languages together to train a multilingual model is a double-edged sword. Although it is simple and  maximizes the sharing across languages, it also brings confusion between languages during recognition.

If the multilingual model can condition on the language identity (LID),  significant improvement can be obtained over the universal multilingual model without LID because the one-hot LID guides the ASR model to generate the transcription of the target language by reducing the confusion from other languages \cite{toshniwal2018multilingual, li2018multi, kannan2019large, zhu2020multilingual, pratap2020massively, sun2021improving}. However, such a multilingual E2E model with LID relies on the prior knowledge of which language the user will speak for every utterance, working more like a monolingual model. To remove the dependency on knowing one-hot LID in advance, one way is to estimate the LID and use it as the additional input to E2E multilingual  models. However, the gain is very limited, especially for streaming E2E models, because the estimation is not very reliable \cite{waters2019leveraging, punjabi2020streaming}. 

\begin{figure}[t]
  \centering
  \includegraphics[width=0.85\linewidth]{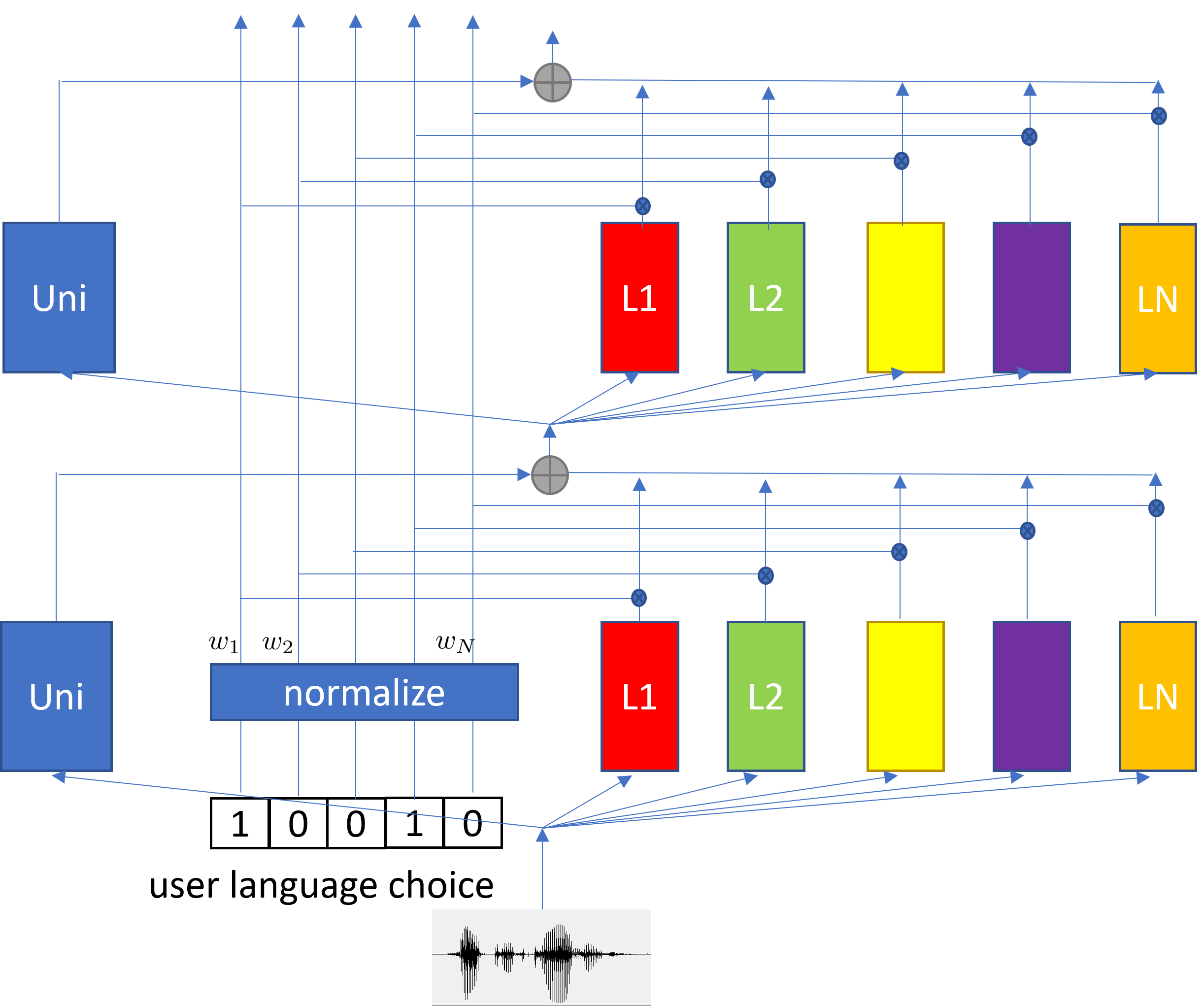}
  \caption{Diagram of configurable multilingual model (CMM). Uni denotes a universal multilingual module, and Li denotes the $i$th language.}
  \label{fig:CMM}
\end{figure}

Because there are more multilingual users than monolingual users  \cite{multilingual}, there is an increasing demand of building multilingual systems which can recognize  multiple languages without knowing in advance which language each individual utterance is from. One solution is to build a corresponding multilingual E2E model for any set of language combinations. However, this is impractical because the number of possible language combinations is huge given so many languages in the world. Zhou et al. proposed a configurable multilingual model (CMM) \cite{zhou2022configurable} which is trained only once by pooling the data from all languages, but can be configured to recognize speeches from any combination of languages. As shown in Figure \ref{fig:CMM}, the hidden output of CMM is calculated as the weighted sum of the output from a universal module and the outputs from all language-specific modules. At runtime, the universal module together with corresponding language-specific modules are activated based on the user choice. Because multilingual users usually speak a few languages, CMM reduces the language confusion space from dozens to a few. Therefore, it performs much better than the multilingual model without LID, and can serve the multilingual users well.

There are several factors to be considered when designing multilingual E2E models. When scaling the multilingual model to cover a large amount of languages (e.g., 50 languages in \cite{pratap2020massively}), the data from those languages is heavily imbalanced, which usually leads to a model performing well on resource-rich languages while failing on low-resource languages. To solve that issue, data sampling is usually used to balance the training data amount \cite{kannan2019large, pratap2020massively}. The model capacity should also be expanded to recognize a large amount of languages. In \cite{pratap2020massively, li2021scaling}, multilingual models are scaled to billions of parameters.  

Because there are differences across languages, it may not be optimal if a simple E2E structure is used to blindly model all the multilingual data. In \cite{kannan2019large}, language-specific adapt layers are used to further boost the multilingual model performance. Similar ideas are used with the mixture of expert structure for multilingual models \cite{das2021multi, gaur2021mixture}.

While E2E models have achieved remarkable success in multilingual modeling, an even more challenging topic is the ASR of code-switching (CS) utterances which contain mixed words or phrases from multiple distinct languages. There are two categories of CS: one is intra-sentential CS where switches happen within an utterance and the other is inter-sentential CS where switches occur between utterances with monolingual words inside each utterance. The former one is  more difficult, attracting most studies.  In \cite{seki2018end}, synthetic  utterance-level CS text was generated to improve the AED model's ASR accuracy on inter-sentential CS utterances. The challenge of intra-sentential CS to E2E models was addressed in \cite{li2019towards} by using the LID model to linearly adjust the posterior of a CTC model, followed by lots of recent studies (e.g., \cite{khassanov2019constrained, yue2019end, lu2020bi, qiu2020towards, nj2020investigation, dalmia2021transformer, zhang2021rnn, diwan2021multilingual}). One major challenge to CS ASR is the lack of the CS data while E2E models are more data hungry. Therefore, methods have been studied to effectively leverage monolingual data. A constrained output embedding method \cite{khassanov2019constrained}  was proposed to encourage the output embedding of monolingual languages similar in order to let the ASR model easily switch between languages. In \cite{lu2020bi, dalmia2021transformer}, a bi-encoder which has a separate encoder for each language is used to train the model from monolingual data. This was further extended with both language-specific encoder and   language-specific decoder to recognize CS speech \cite{zhou2020multi}.

\section{Adaptation}
\label{sec:adaptation}
The performance of ASR systems can  degrade significantly when the test conditions differ from training. Adaptation algorithms are designed to address such challenging issues. A comprehensive overview of adaptation technologies in ASR can be found in \cite{bell2020adaptation}. In this section, we highlight some adaptation technologies for E2E models. 

\subsection{Speaker Adaptation}
Speaker adaptation adapts ASR models to better recognize a target speaker's speech. It is a common practice to adapt the acoustic encoder of CTC \cite{li2018speaker, do2021multiple}, AED \cite{weninger2019listen, meng2019speaker}, RNN-T \cite{sim2019personalization, huang2020rapid} and Conformer Transducer \cite{huang2021rapid}. Another popular methodology is to augment the input speech with speaker embeddings \cite{delcroix2018auxiliary, sari2020unsupervised, zhao2020speech}.

The biggest challenge of speaker adaptation is that the adaptation data amount from the target speaker is usually very small. There are several ways to address such a challenge. 
The first one is  using regularization techniques such as Kullback-Leibler (KL) divergence regularization \cite{yu2013kl}, maximum a posteriori adaptation \cite{huang2015maximum}, or  elastic weight consolidation \cite{kirkpatrick2017overcoming}, to ensure the adapted models do not overfit the limited adaptation data \cite{li2018speaker, sim2019personalization, weninger2019listen, meng2019speaker}. 
The second approach is multi-task learning. E2E models typically use subword units as the output target in order to achieve high recognition accuracy. Sometimes, the number of subword units is at the scale of thousands or even more. These units usually cannot be fully observed in the limited speaker adaptation data. In contrast, the small number of character units usually can be fully covered. Therefore, multi-task learning using both character and subword units can significantly alleviate such sparseness issues \cite{li2018speaker, meng2019speaker}.
The third approach is to use multi-speaker text-to-speech (TTS) to expand the adaption set for the target speaker \cite{sim2019personalization, huang2020rapid}. Because the transcription used to generate the synthesized speech is also used for model adaptation, this approach also alleviates the hypothesis error issue in unsupervised adaptation when combining the synthesized speech with the original speech from the target speaker.

It is desirable to use E2E models on devices because of the compact model size. There are interesting works \cite{sim2019investigation, sim2021robust} which study how to do secure adaptation on devices with continuous learning so that the user data never leaves devices.  Several engineering practices were described in order to deal with the limited memory and computation power on devices.

\subsection{Domain Adaptation}
\label{ssec:domain_adapt}
Domain adaptation is the task of adapting ASR models to the target domain which has content mismatch from the source domain in which the ASR models were trained. Because E2E models tend to memorize the training data well, their performance usually degrades a lot in a new domain. Such a challenge attracts much more works on domain adaptation than speaker adaptation. This is not a severe issue for hybrid models because their language model (LM) is trained with a much larger amount of text data than that is used in the paired speech-text setup for E2E model training. 
The big challenge of adapting E2E models to a new domain is that it is not easy  to get enough paired speech-text data in the new domain, which requires transcribing the new-domain speech. However, it is much easier to get text data in the new domain. Therefore, the mainstream of domain adaptation methods for E2E models focuses on the new-domain text only.  

A widely adopted approach to adapt E2E models to a new domain is fusing E2E models with an external LM trained with the new-domain text data. There are several LM fusion methods, such as shallow fusion \cite{gulcehre2015Shallow}, deep fusion  \cite{gulcehre2015Shallow}, and cold fusion \cite{Sriram2018Cold}. Among them, the simplest and most popular method is shallow fusion in which the external LM is log-linearly interpolated with the E2E model at inference time \cite{toshniwal2018comparison}. 

However, shallow fusion does not have a clear probabilistic interpretation. As an improvement, a maximum a posteriori based decoding method \cite{kanda2016maximum, kanda2017maximum} was proposed for the integration of an external LM for CTC. A density ratio approach based on Bayes' rule was proposed in \cite{mcdermott2019density} for RNN-T. During inference, the output of the E2E model is modified by the ratio of the external LM trained with the new-domain text data and the source-domain LM trained with the transcript of the training set which has the paired speech-text data for E2E model training. Another similar model is the hybrid autoregressive transducer (HAT) model \cite{variani2020hybrid} designed to improve the RNN-T model. In the HAT model, the label distribution is derived by normalizing the score functions across all labels excluding blank. Hence, it is mathematically justified to integrate the HAT model with an external LM using the density ratio method. 

In \cite{meng2021internal}, an internal LM estimation (ILME)-based fusion was proposed to enable a more effective LM integration. During inference, the internal LM score of an E2E model is estimated by zeroing out the contribution of the encoder and is subtracted from the log-linear interpolation between E2E and external LM scores. An internal LM training (ILMT) method  \cite{meng2021ILMT} was further proposed to minimize an additional internal LM loss by updating only the components that estimate the internal LM. ILMT increases the modularity of the E2E model and alleviates the mismatch between training and ILME-based fusion. 

Tuning the LM weights on multiple development sets is computationally expensive and time-consuming. To eliminate the weights tuning, the MWER training with LM fusion was proposed in \cite{peyser2020improving, meng2021minimum} where the LM fusion is performed during MWER training. During inference, LM weights pre-set in training enables a robust LM fusion on test sets from different domains.

Because LM fusion methods require interpolating with an external LM, both the computational cost and footprint are increased, which may not be applicable to ASR on devices. With the advance of TTS technologies, a new trend is to adapt E2E models with the synthesized speech generated from the new-domain text \cite{sim2019personalization, peyser2019improving, Li2020Developing, zheng2021using}. This is especially useful for adapting RNN-T, in which the prediction network works similarly to an LM. It was shown that such domain adaptation method with TTS is more effective than LM fusion methods \cite{Li2020Developing}. 

\begin{figure}[t]
  \centering
  \includegraphics[width=1.0\linewidth]{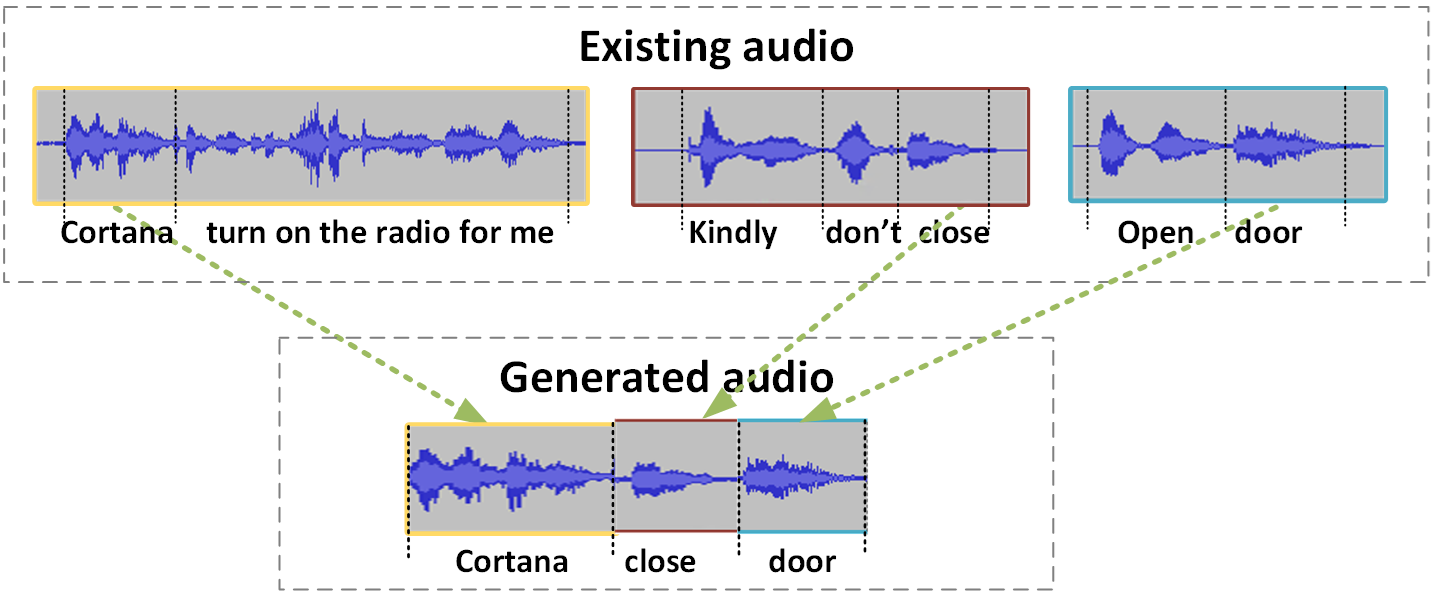}
  \caption{An example of the spliced data method. It generates ``Cortana close door'' from 3 utterances from the source training data by extracting the audio segments of ``Cortana'', ``close'', and ``door'' out and then concatenating them.}
  \label{fig:splicing}
\end{figure}

The TTS-based adaptation method also has its drawbacks. The first is that TTS speech is different from the real speech. It sometimes also degrades the recognition accuracy on real speech \cite{li2019semi}. This issue was alleviated in \cite{kurata2021improving} by inserting a mapping network before the encoder network when the input is TTS audio in both the training and adaption stage. This mapping network works as feature space transformation to map the space of TTS audio to the space of real speech. During testing with real speech, the mapping network was removed. The second is that the speaker variation in the TTS data is far less than that in the large-scale ASR training data. The third is the cost of training a multi-speaker TTS model and the generation of synthesized speech from the model is large. These issues were solved by a spliced data method \cite{zhao2021addressing} which was used to adapt E2E models with better performance than the TTS-based adaptation method. For any text sentence in the new domain, the method  randomly extracts the corresponding audio segments from the source training data, and then concatenates them to form new utterances. Figure \ref{fig:splicing} gives an example of how this spliced data method works.  Similar ideas have been used for data augmentation by replacing some word segments in an utterance with new word segments from another utterance to train a general E2E model \cite{sun2021semantic, lam2021fly}.

\begin{figure}[t]
  \centering
  \includegraphics[width=1.0\linewidth]{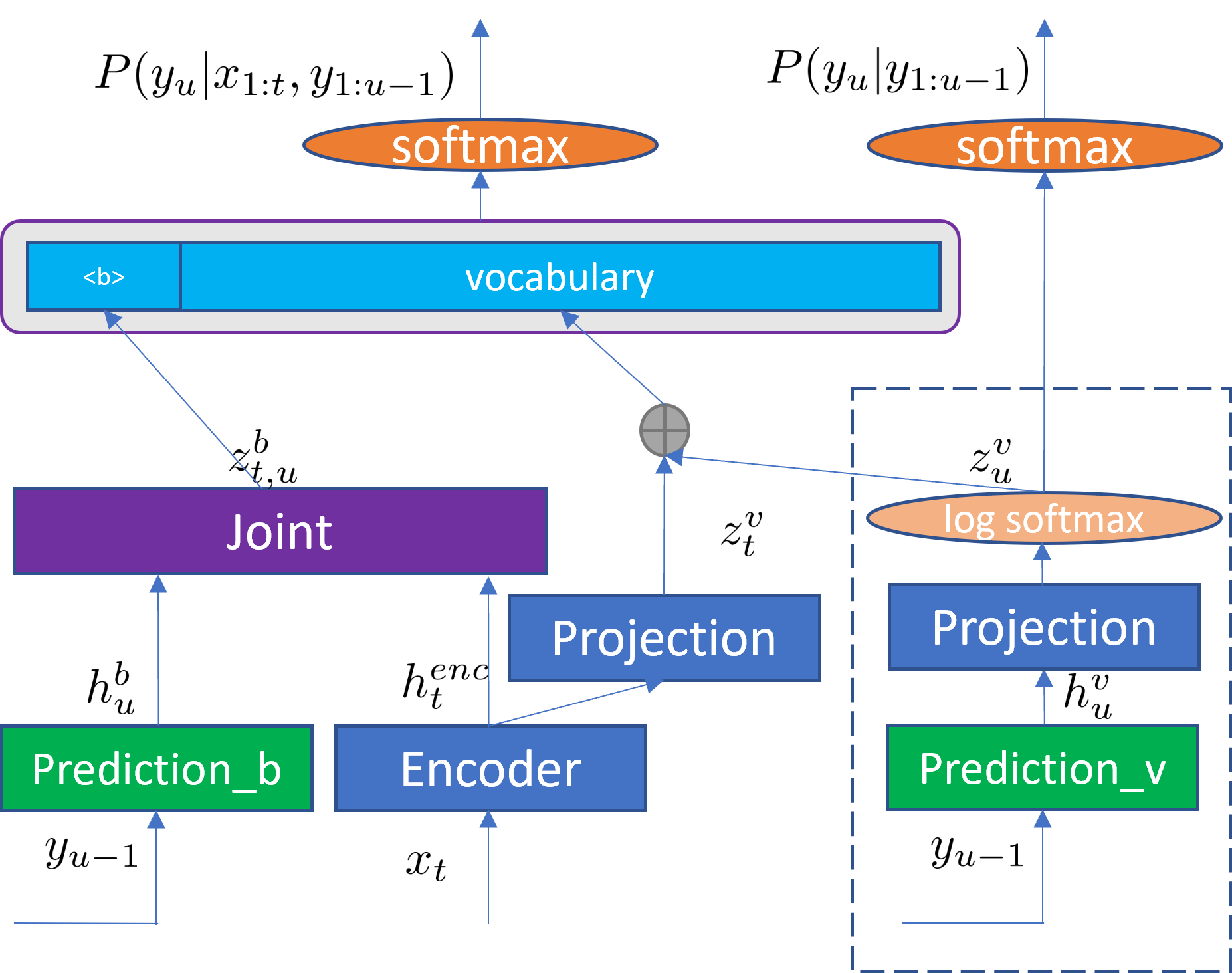}
  \caption{Flowchart of factorized neural transducer model which separates the prediction of blank and normal tokens in the vocabulary. The superscript $v$ and $b$ denote the vocabulary and blank, respectively.}
  \label{fig:fnt}
\end{figure}

Both the TTS-based adaptation and the spliced data method need to synthesize audios from text, and then update E2E models. Recently, a fast text adaptation method \cite{pylkkonen2021fast} was proposed to adapt the prediction network of RNN-T by treating it as an LM and then using the text from the new domain to update it. In \cite{meng2021internal2}, ILMT \cite{meng2021ILMT} was used to ensure that the internal LM inside RNN-T behaves similarly to a standalone neural LM, and then it can be adapted with text only data. 

However, as shown in \cite{ghodsi2020rnn}, the prediction network in RNN-T does not fully function as an LM. Chen et al. showed this is because the prediction network needs to predict both normal tokens and blank, and therefore proposed a factorized neural transducer which separates the prediction of normal tokens and blank \cite{chen2021factorized}. As shown in Figure \ref{fig:fnt}, the modules in the dashed box function exactly as an LM, predicting the probability $P(y_u | {\bf{y}}_{1:u-1})$. Therefore, it can be updated with the text from the new domain by using any well-established neural LM adaptation methods. Finally, the adapted factorized neural transducer with the branch output $P(y_u |{\bf{x}}_{1:t}, {\bf{y}}_{1:u-1})$ is used to recognize the speeches from the new domain with large accuracy improvement \cite{chen2021factorized}. 

While these domain adaptation technologies are powerful, it is desirable that the adaptation of E2E models to the new domain does not degrade the models' performance on the source domain. Therefore, it is worth looking at the concept of lifelong learning which enables E2E models to learn new tasks without forgetting the previously learned knowledge \cite{chang2021towards}. If the source-domain data is available, it can be mixed with the new-domain data for adaptation \cite{zhao2021addressing}. Otherwise, the popular solution is to use regularization that prevents the adapted model from moving too far away from the source model \cite{meng2021internal2}. 

\subsection{Customization}
\label{ssec:customization}
Different from speaker adaptation which adjusts a speaker independent E2E model toward the speaker audio and domain adaptation which lets an E2E model recognize the content of a new domain better, customization here specifically refers to the technologies that leverage context such as contacts, location, music playlist etc., of a specific user to significantly boost the ASR accuracy for this user. For example, an English ASR system usually cannot recognize the contact names of a Chinese person well. However, if the English ASR system is presented with the contact list of this Chinese person, the ASR output can be biased toward the contact names.  Such biasing is even more effective when designed with context activation phrases such as "call", "email", "text", etc.

\begin{figure}[t]
  \centering
  \includegraphics[width=0.85\linewidth]{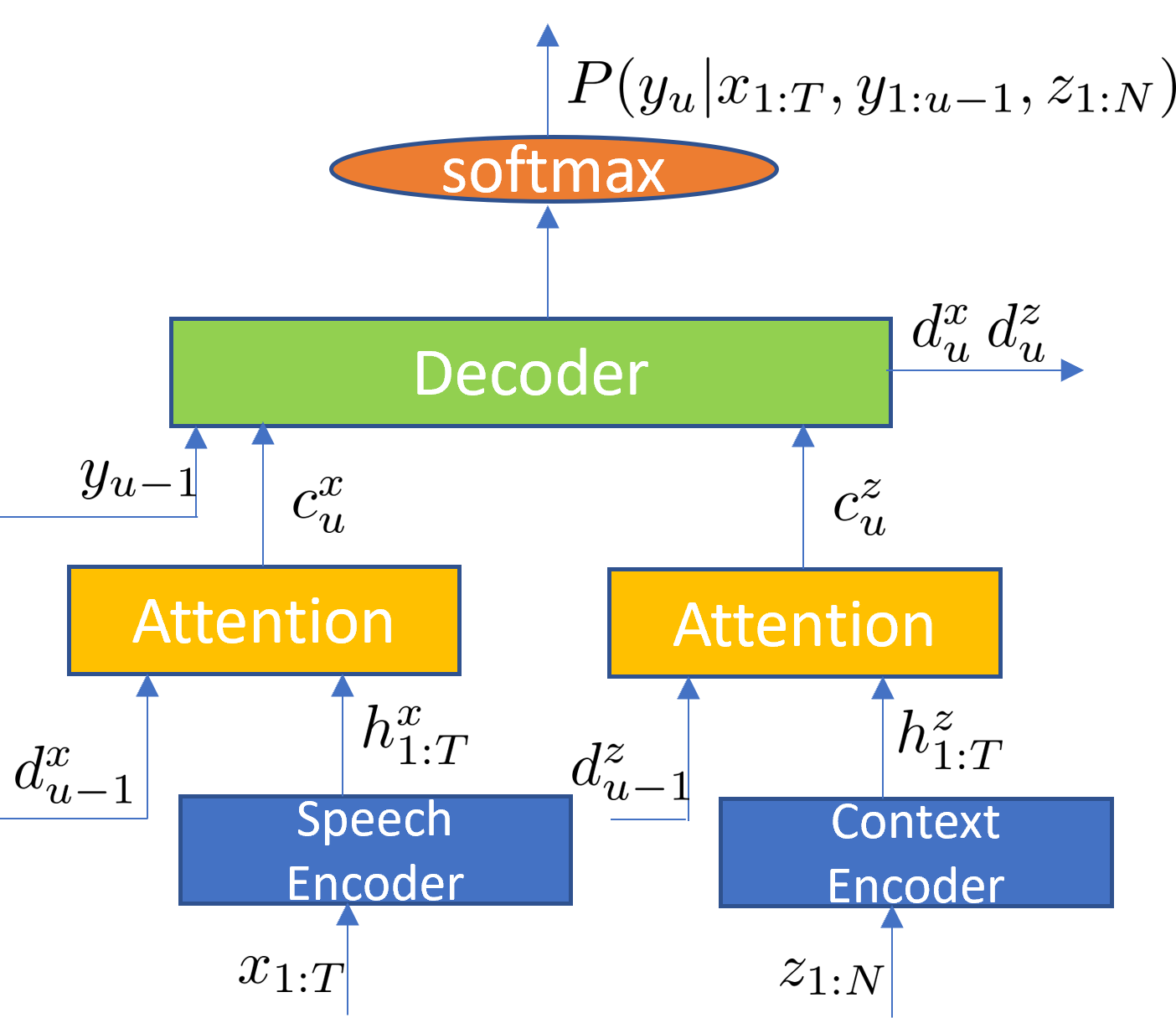}
  \caption{Flowchart of Contextual LAS \cite{pundak2018deep}}
  \label{fig:CLAS}
\end{figure}

While it is relatively easier for hybrid systems to do so with the on-the-fly rescoring strategy which dynamically adjusts the LM weights of a small number of n-grams which are relevant to the particular recognition context, it is more challenging to E2E systems. One solution is to add a context bias encoder in addition to the original audio encoder into the E2E model, which was first proposed in \cite{he2017streaming} to bias a keyword spotting E2E model toward a specific keyword. The idea was then extended into a contextual LAS (CLAS) ASR model \cite{pundak2018deep}, plotted in figure \ref{fig:CLAS}. LAS (Listen, Attend and Spell) \cite{chan2016listen} is one particular instantiation of AED. Compared to the standard AED model in figure \ref{fig:AED}, CLAS adds the context encoder in the bottom right of the figure which takes the context list ${\bf{z}}_{1:N}$ as the input, where ${\bf{z}}_i$ denotes the $i$-th phrase.  An attention module is used to generate the contextual vector ${\bf{c}}_u^z$ as one of the inputs to the decoder. In this way, the network output conditions not only on the speech signal and previous label sequence but also on the contextual phrase list. It was further extended into a two-step memory enhanced model in \cite{huber2021instant}. It is common that a contextual phrase list contains rare words, especially names which are not observed during training. In such a case, E2E models have not learned to map the rare words' acoustic signal to words. This issue can be alleviated by further adding a phoneme encoder in addition to the text encoder for the contextual phrase list \cite{bruguier2019phoebe, chen2019joint}. The same idea has also been applied to RNN-T \cite{jain2020contextual}. 

However, as shown in \cite{pundak2018deep}, it becomes challenging for the bias attention module to focus if the  biasing list is too large (e.g., more than 1000 phrases). Therefore, a more popular way to handle a large contextual phrase list is shallow fusion with the contextual biasing LM \cite{zhao2019shallow, le2021deep}. 
A challenge for the fusion-based biasing method is that it usually benefits from prefix which may not be available all the time.  In \cite{huang2020class}, class tags are inserted into word transcription during training to enable context aware training. During inference, class tags are used to construct contextual bias finite-state transducer. In \cite{le2021contextualized}, trie-based deep biasing, WFST shallow fusion and neural network LM contextualization are combined together to reach good biasing performance for tasks with and without prefix. 
As discussed in Section \ref{sec:adaptation}.\ref{ssec:domain_adapt}, density ratio and HAT perform better than shallow fusion when integrating an external LM. Contextual biasing with density ratio and HAT   were also shown to have better biasing performance \cite{andres2021contextual, sainath2021efficient}. 

\begin{figure}[t]
  \centering
  \includegraphics[width=0.85\linewidth]{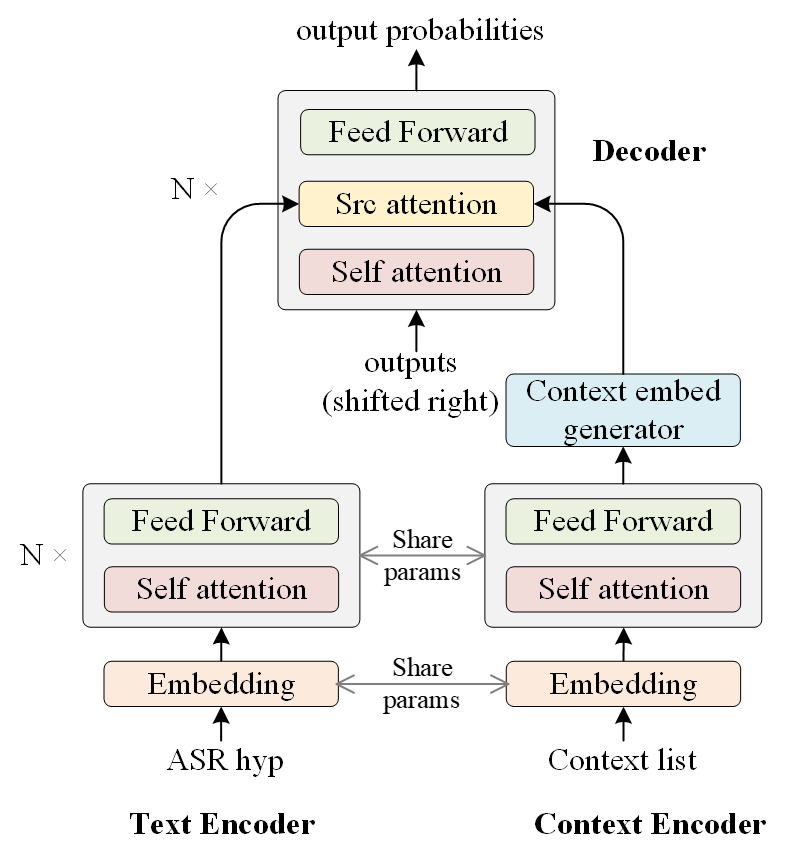}
  \caption{Flowchart of Context Spelling Correction}
  \label{fig:CSC}
\end{figure}

The most challenging part of context biasing is how to deal with a very large bias list which contains more than 1000 phrases. Wang et al. \cite{wang2021a} provided a nice solution by extending spelling correction \cite{guo2019spelling} to contextual spelling correction (CSC), which is shown in Figure \ref{fig:CSC}. Both the embeddings of the ASR hypothesis and the contextual phrase list are used as the input to the decoder, which generates a new word sequence. A filtering mechanism based on the distance between ASR hypothesis and contextual phrases is used to trim the very large phrase list to a relatively small one so that the attention can perform well. Such filtering is the key to the success which cannot be done inside ASR models such as CLAS because the filtering relies on the ASR decoding results. 

\section{Advanced Models}

In this section, we discuss several advanced models.

\label{sec:models}

\subsection{Non-Autoregressive Models}
While most E2E models use autoregressive (AR) modeling to predict target tokens in a left-to-right manner as Eqs. \eqref{eq:Prob_attention}, there is a recent trend of using non-autoregressive (NAR) modeling which generates all target tokens simultaneously with one-shot or iteratively without replying on predicted tokens in early steps \cite{higuchi2020mask, bai2020listen, tian2020spike, chen2020non, song2021non, fan2021cass, higuchi2021improved}. These NAR methods root from the assumption that the feature sequence generated by the acoustic encoder contains not only acoustic information but also some language semantic information. However, such an assumption is not very strong, resulting in worse performance of NAR modeling compared to AR modeling in general. The biggest advantage of NAR models is that its decoding speed is much faster than that of AR models because there is no dependency on previous tokens. All target tokens can be predicted in parallel while the decoding of AR models is more complicated because of the token dependency. 

A typical way to generate all target tokens is described in \cite{bai2020listen}. The target token sequence length $L$ is predicted or set as a constant value. Then the NAR model assumes that each token is independent of each other as 
\begin{equation} \label{eq:Prob_nar} 
P( {\bf{y}|\bf{x}} ) = \prod_{u=1}^L P(y_u | {\bf{x}}).
\end{equation}
At decoding time, the predicted token at each position is the one with the highest probability.

\begin{figure}[t]
  \centering
  \includegraphics[width=0.75\linewidth]{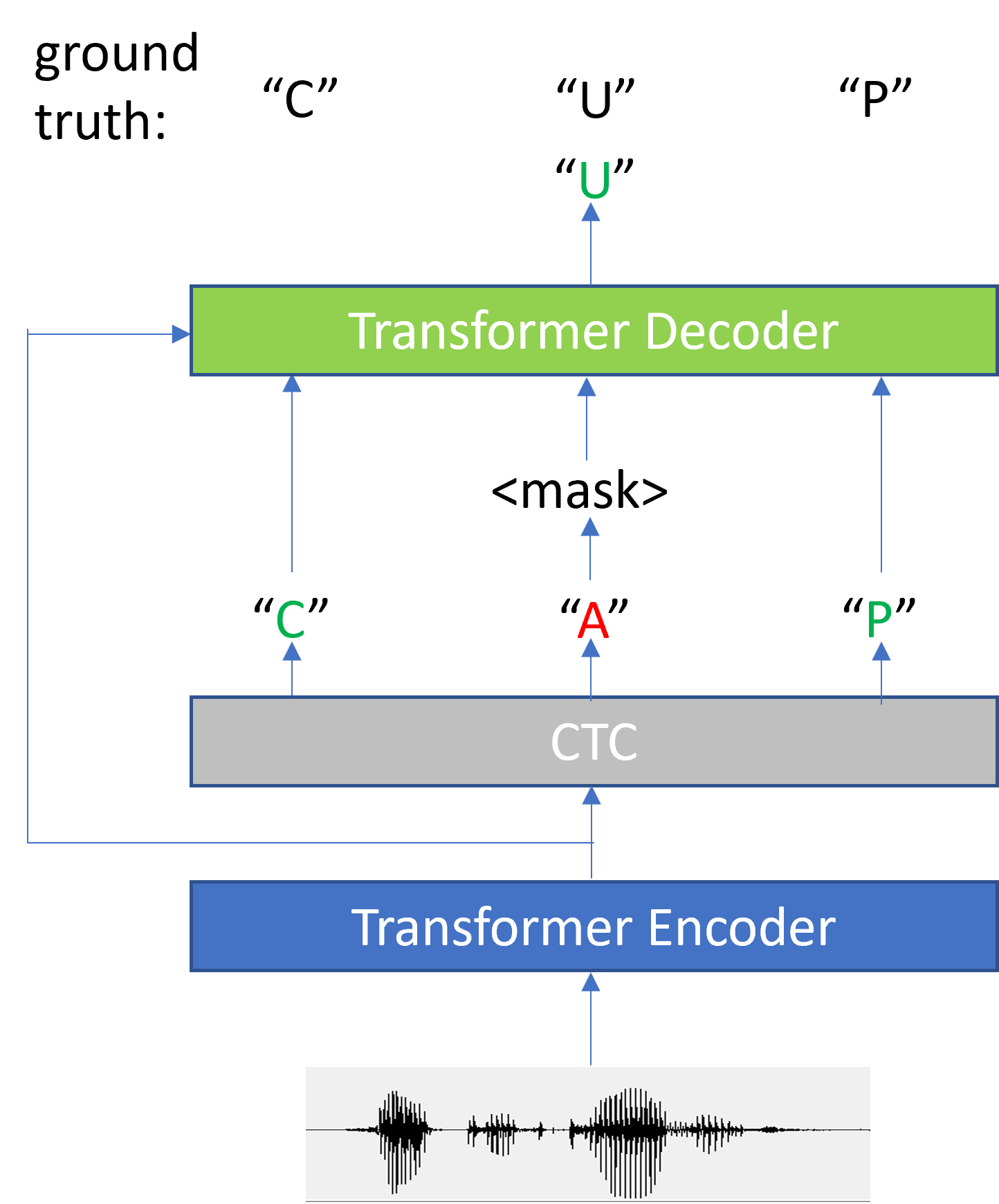}
  \caption{An example of Mask CTC \cite{higuchi2020mask}. The CTC model generates a token sequence ``C A T'' in which ``A'' has a low confidence score. Then ``A'' is masked. The speech sequence and unmasked tokens are used to predict the token ``U''.}
  \label{fig:MaskCTC}
\end{figure}

The independence assumption in Eq. \eqref{eq:Prob_nar} is very strong. The token sequence $L$ is also hard to predict usually. Mask CTC \cite{higuchi2020mask} was proposed to solve these issues. It predicts a set of masked tokens ${\bf{y}}_{mask}$, conditioning on the observed tokens ${\bf{y}}_{obs} = {\bf{y}} \setminus {\bf{y}}_{mask}$ and the input speech sequence ${\bf{x}}$ as
\begin{equation} \label{eq:Prob_maskCTC} 
P( {\bf{y}}_{mask}|{\bf{y}}_{obs}, {\bf{x}} ) = \prod_{y \in {\bf{y}}_{mask} } P(y | {\bf{y}}_{obs}, {\bf{x}}).
\end{equation}
Figure \ref{fig:MaskCTC} shows  an example of how Mask CTC works. The target sequence was first initialized with the CTC outputs. Then tokens with low confidence scores are masked and are iteratively optimized conditioning on the unmasked tokens and input speech sequence. The length of the final token sequence generated by Mask CTC is the same as the length of the CTC initial output, therefore Mask CTC can only deal with the substitution error. In \cite{higuchi2021improved}, the length of a partial target sequence is predicted to enable Mask CTC's ability of handling deletion and insertion errors. 

\subsection{Unified Models}

The most significant component in E2E models is the encoder, which is well studied in literature. There are lots of encoder designs depending on the requirements of  streaming, latency, and computational cost in different target scenarios. The development cost is formidable if we train a separate E2E model for each application scenario. Therefore, an emerging trend is to train a single unified model which can be configured with multiple streaming, latency, and computational cost setups during decoding. In \cite{yu2021dual}, a dual E2E model was proposed with shared weights for both streaming and non-streaming ASR. In order to have a dual-mode Conformer encoder, the authors designed dual-mode modules for the components such as convolution, pooling, and attention layers inside Conformer. An in-place T/S learning was used to train the student streaming mode from the full-context non-streaming mode. Such training even brought accuracy and latency benefits to the streaming ASR. 
Another model unifying streaming and non-streaming modes was proposed in \cite{audhkhasi2021mixture} which decomposes the Transformer's softmax attention into left-only causal attention and right-only anti-causal attention. For the streaming scenario, it just uses the causal attention while the non-streaming mode uses both attentions.

\begin{figure}
     \centering
     \begin{subfigure}[b]{0.15\textwidth}
         \centering
         \includegraphics[width=\textwidth]{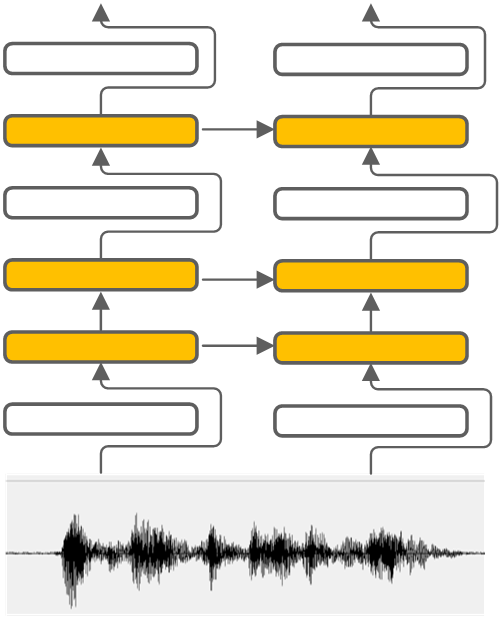}
         \caption{training with layer dropout}
         \label{fig:dyn_training}
     \end{subfigure}
     \hfill
     \begin{subfigure}[b]{0.15\textwidth}
         \centering
         \includegraphics[width=\textwidth]{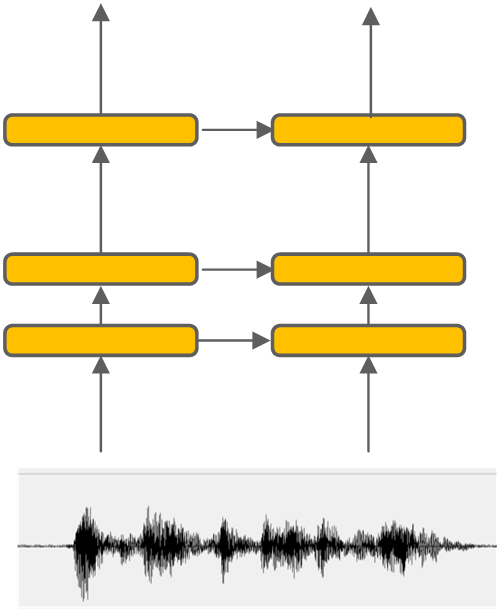}
         \caption{pruned encoder in decoding}
         \label{fig:dyn_prune}
     \end{subfigure}
     \hfill
     \begin{subfigure}[b]{0.15\textwidth}
         \centering
         \includegraphics[width=\textwidth]{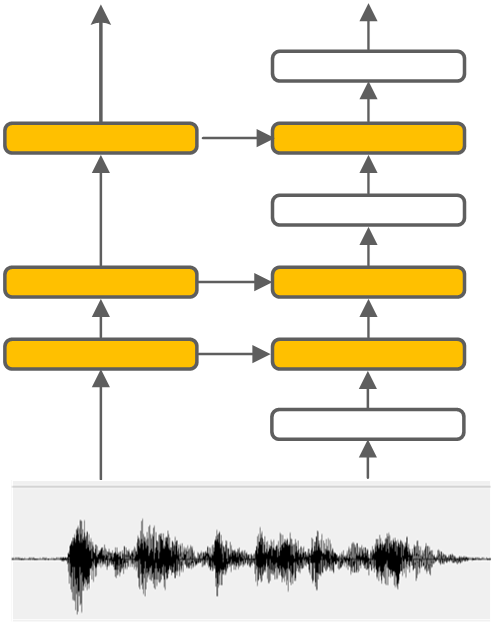}
         \caption{dynamic encoder in decoding}
         \label{fig:dyn}
     \end{subfigure}
        \caption{The training and decoding of dynamic encoder \cite{shi2021dynamic}}
        \label{fig:dynamic_encoder}
\end{figure}

There are several studies working on a unified model with dynamic computational cost during inference. In \cite{wu2021dynamic}, a dynamic sparsity network is used as the encoder of RNN-T. It only needs to be trained once and then can be set with any predefined sparsity configuration at runtime in order to meet various computational cost requirements on different devices. An even more direct way to have a dynamic encoder was proposed in \cite{shi2021dynamic}. During training, layer dropout is applied to randomly drop out encoder layer as shown in Figure \ref{fig:dyn_training}. The pruned model can be used to decode full utterances for constant low computational cost as shown in Figure \ref{fig:dyn_prune}. An example of advanced usage of the dynamic encoder is plotted in Figure \ref{fig:dyn}, where the dynamic encoder is configured with a small number of layers in the beginning of the utterance, and then configured with full layers in the remaining. The idea of having different computational costs within an utterance was also studied in \cite{macoskey2021amortized} for RNN-T. The model has a fast encoder and a slow encoder. An arbitrator is used to select which encoder should be used given an input speech frame. 

It requires a small latency for applications such as voice search and command control, while the applications such as dictation and video transcription usually can afford a larger latency.  Variable context training \cite{tripathi2020transformer, mahadeokar2021flexi, kim2021multi} was proposed to build a unified model which can be configured for different latency requirements at runtime. During training, the Transformer encoder is provided with different right context lengths which correspond to different theoretic latency. In \cite{mahadeokar2021flexi}, the alignment constraint method \cite{mahadeokar2021alignment} was also used to flexibly set latency thresholds for different tasks. A task ID is used at runtime to configure the Transducer model with different encoder segments and latency. 

\subsection{Two-pass Models}

Although a single E2E model can already achieve very good ASR performance, its performance can be further improved with a second-pass model. Spelling correction methods \cite{guo2019spelling, Zhang2019} were proposed by using TTS data to train a separate translation model which is used to correct the hypothesis errors made by the first-pass E2E model. The spelling correction model is a pure text-to-text model without using the speech input. As E2E models need to be trained with paired speech-text data, the language modeling power is always a concern of E2E models. Wang et al. proposed a two-pass RNN-T model in which the first-pass  RNN-T transcribes speech into syllables while the second-pass RNN-T converts syllable sequences into character sequences \cite{wang2021cascade}. Because the second-pass RNN-T uses text instead of speech as input, therefore it can leverage a much larger amount of text data to build a more powerful  capability for language modeling. 

\begin{figure}[t]
  \centering
  \includegraphics[width=0.75\linewidth]{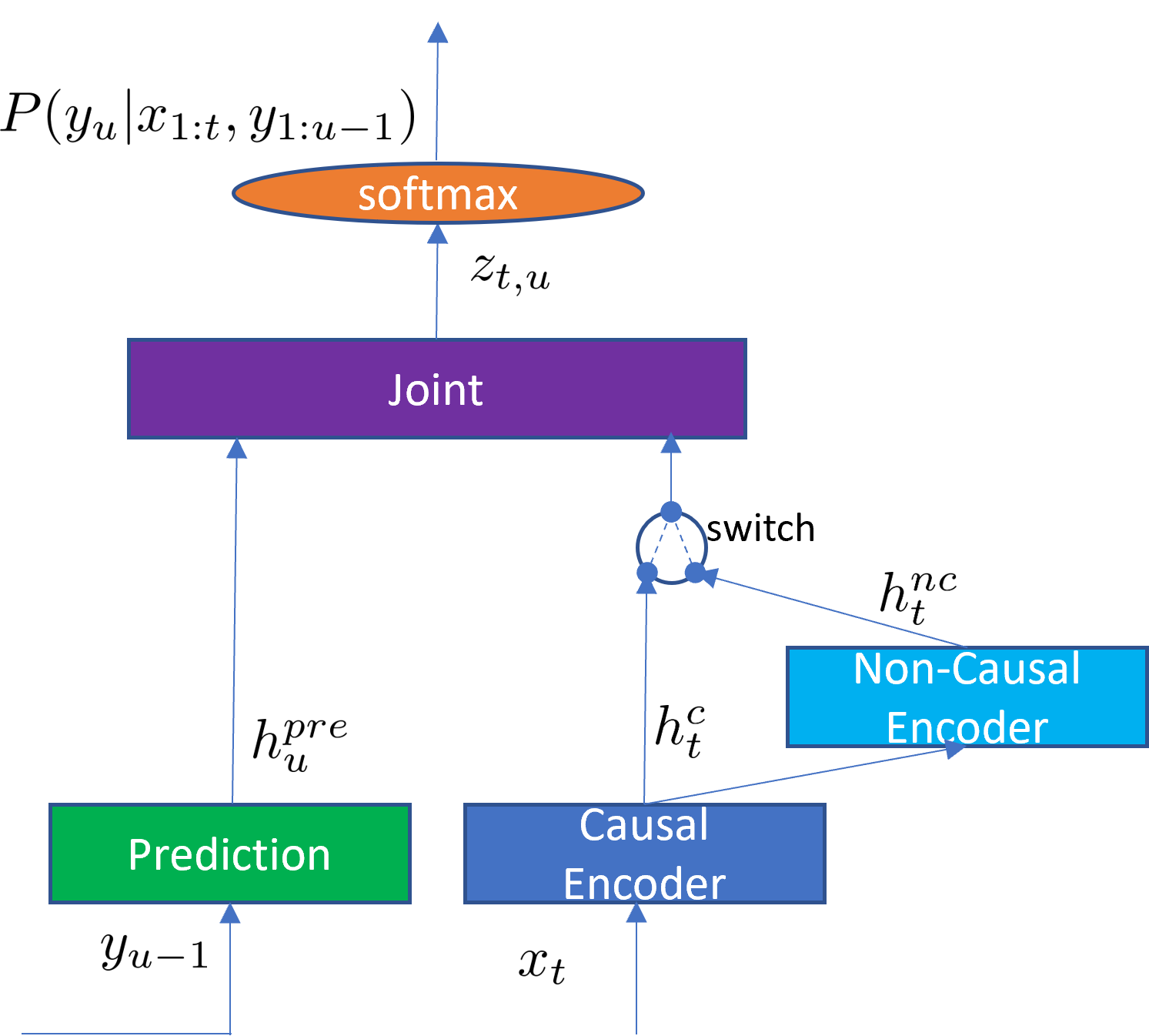}
  \caption{An RNN-T model with a cascaded encoder.}
  \label{fig:cascaded}
\end{figure}

The second-pass processing with decoding hypothesis only cannot leverage the speech input. In \cite{sainath2019two}, a two-pass model was proposed to use an AED decoder to attend the encoder output of a streaming RNN-T model. In this way, the first-pass RNN-T model provides immediate recognition results while the second-pass AED model can provide better accuracy with small additional perceived latency. This work was extended as the deliberation model which uses an LSTM AED decoder  \cite{hu2020deliberation} or a Transformer AED decoder \cite{hu2021transformer} to attend both the encoder output and first-pass decoding hypothesis. 

However, AED models cannot perform well on long utterances \cite{chiu2019comparison}. Therefore, an RNN-T model with the cascaded encoder was proposed \cite{narayanan2021cascaded} as shown in Figure \ref{fig:cascaded}. Such a model can also be considered as a unified model which provides both streaming and non-streaming solutions, and is a special case of Y-model \cite{tripathi2020transformer} which has both streaming and non-streaming branches in a Y-shape. The causal encoder output $h_t^c$ is fed into a non-causal encoder to generate $h_t^{nc}$. Depending on applications, a switch is used to let $h_t^c$ or $h_t^{nc}$ go to the joint network.  In the context of two-pass modeling, the causal output $h_t^c$ is used in the first pass and the non-causal output $h_t^{nc}$ is used in the second pass. Such a cascaded model is trained in one stage, while the first-pass and second-pass models in deliberation are trained in two stages. Another advantage of the RNN-T model with a cascaded encoder is its better performance on long utterances because the RNN-T decoder  better handles long utterances than the AED decoder. 

Note that the second-pass model brings additional latency and computational costs to ASR systems, therefore, careful designs have been studied to hide these costs in commercial systems \cite{chang2020low, li2021better, sainath2021efficient}.

\subsection{Multi-talker Models}
\label{ssec:mtalker}
While ASR systems have achieved very high recognition accuracy in most single-speaker applications, it is still very difficult to achieve satisfactory recognition accuracy in scenarios with multiple speakers talking at the same time. A common practice in the industry is to separate the overlapped speech first and then use an ASR model to recognize the separated speech streams \cite{yoshioka2019advances}. The biggest challenge to multi-talker overlapping ASR is the permutation problem which occurs when the mixing sources are symmetric and the model cannot predetermine the target signal for its outputs. Deep clustering \cite{hershey2016deep} and permutation invariant training (PIT) \cite{yu2017permutation} were proposed to address such a challenge to separate overlapping speech signals. Specifically, PIT is simpler to implement and easier to be integrated with other methods. Therefore, it becomes the most popular speech separation method. Instead of the two-stage processing, Yu et al. \cite{yu2017recognizing} proposed to directly optimize the ASR criterion with a single model using PIT without having an explicit speech separation step.

\begin{figure}
     \centering
     \begin{subfigure}[b]{0.5\textwidth}
         \centering
         \includegraphics[width=\textwidth]{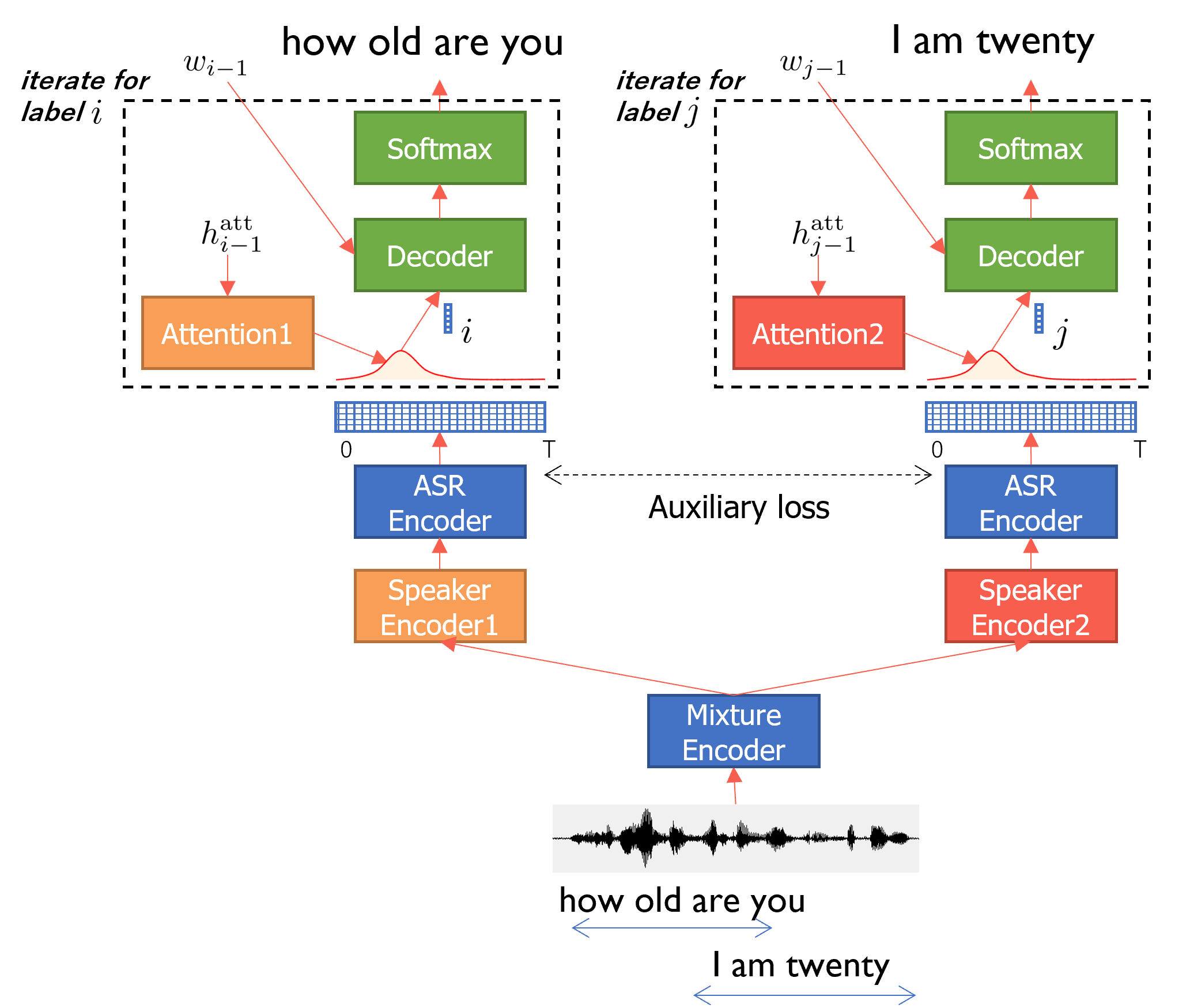}
         \caption{multi-talker AED model with PIT}
         \label{fig:PIT}
     \end{subfigure}
     \hfill
     \begin{subfigure}[b]{0.3\textwidth}
         \centering
         \includegraphics[width=\textwidth]{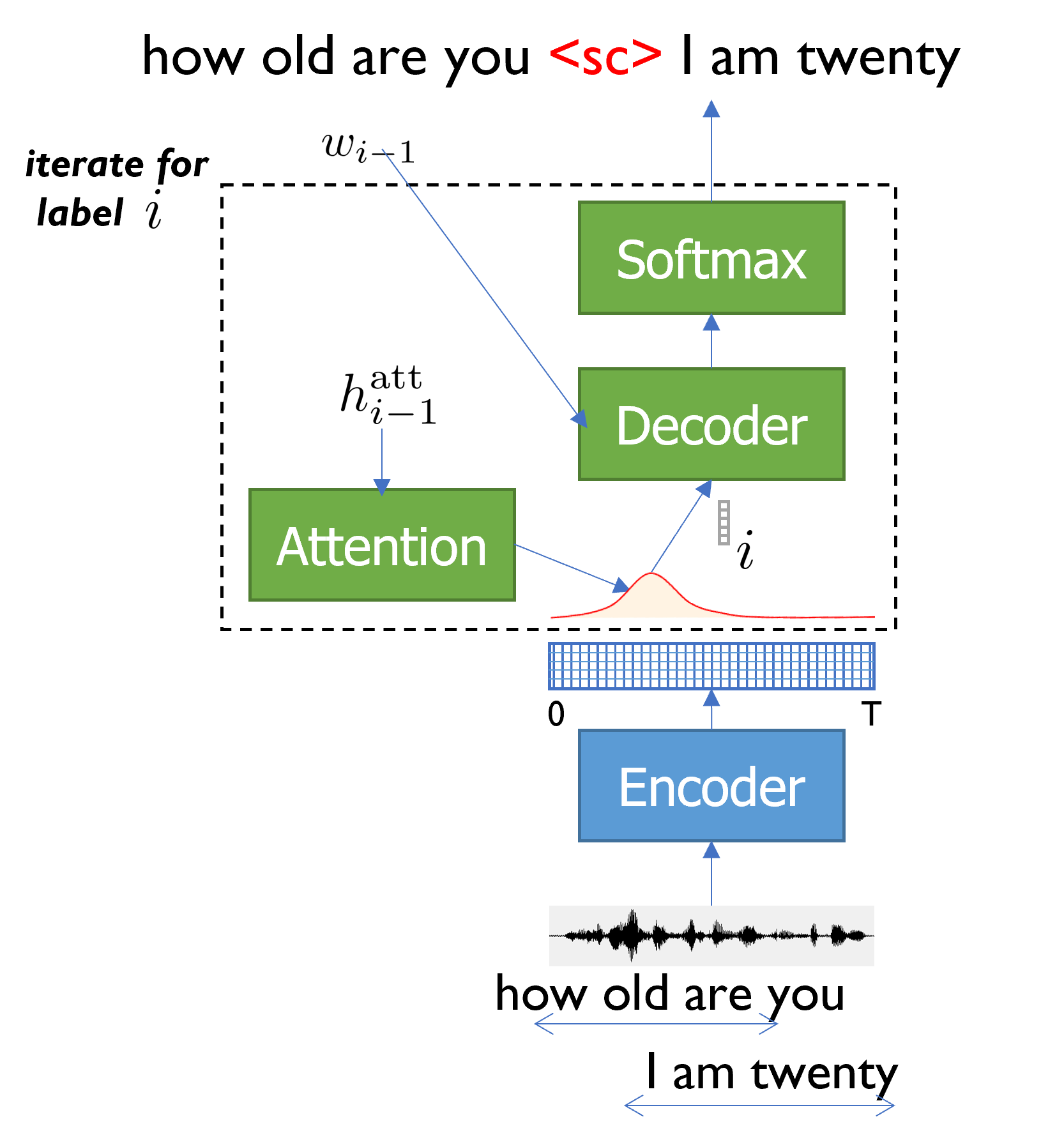}
         \caption{multi-talker AED model with SOT. $\langle sc \rangle$ is a symbol representing speaker change inserted between label sequences of speakers.}
         \label{fig:SOT}
     \end{subfigure}
     \hfill
     \begin{subfigure}[b]{0.5\textwidth}
         \centering
         \includegraphics[width=\textwidth]{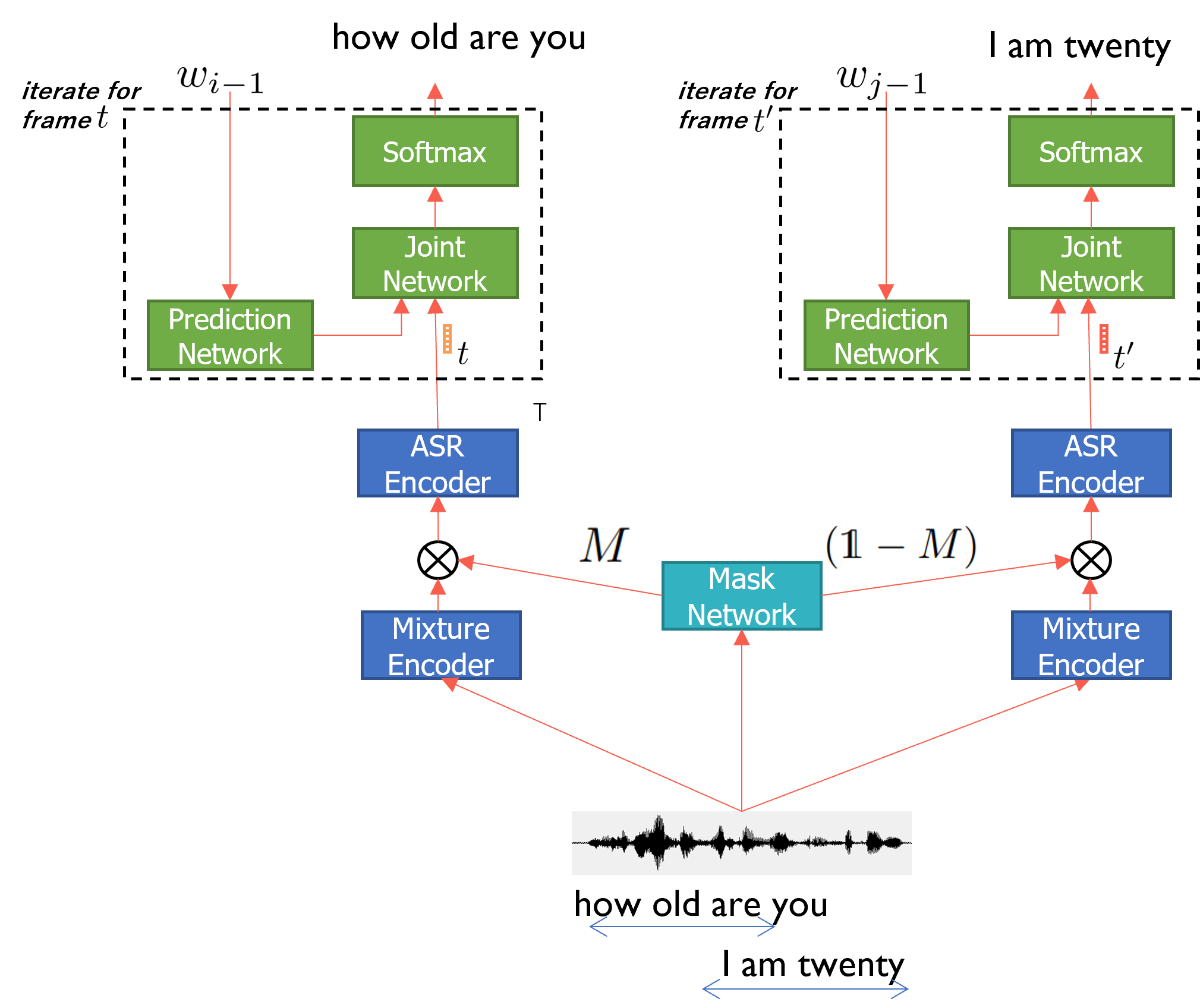}
         \caption{streaming multi-talker RNN-T model with HEAT}
         \label{fig:HEAT}
     \end{subfigure}
        \caption{Architecture of E2E models to recognize overlapping speech}
        \label{fig:multi-talker}
\end{figure}

The first E2E model for overlapping speech was proposed in \cite{settle2018end} with PIT loss in the label level without using the source signal from each speaker. It was further extended in \cite{chang2019end} without pretraining, in \cite{chang2019mimo} for multi-channel input and multi-channel output, and in \cite{chang2020end} with Transformer. Figure \ref{fig:PIT} shows the network architecture of these methods. The overlapping speech goes through a mixture encoder for separation followed by two branches which generate the recognition results from two speakers. Most modules are shared between two branches except the attention module and speaker encoder. For the overlapping speech with $S$ speakers, the training loss is calculated by considering all possible speaker permutation $\Phi(1,...,S)$ as
\begin{equation}
    \label{eq:pitasr}
    L^{PIT} = \min_{\phi \in \Phi(1,...,S)} \sum_{s=1}^S CE( {\bf{y}}^s, {\bf{r}}^{\phi[s]}),
\end{equation}
where $CE()$ is the cross entropy function with the $s$-th output ${\bf{y}}^s$ and the permuted reference ${\bf{r}}^{\phi[s]}$.

There are three theoretical limitations in the models with PIT-ASR loss in Eq \eqref{eq:pitasr}. First, the number of output branches is the same as  $S$, the number of the speakers. After the model is trained, it cannot handle the overlapping speech with more speakers. Second, the training cost is $O(S^3)$ using the Hungarian algorithm, which prevents it from being applied to scenarios with a large number of speakers. Third, there is possible leakage with duplicated decoding hypothesis between output branches because the outputs in different branches do not have a direct dependency.  To address these limitations, Kanda et al. proposed a simple but effective serialized output training (SOT) method \cite{kanda2020serialized} which uses a single AED model to predict the merged label sequence $\Psi(1,...,S)$ from all speakers as in Figure \ref{fig:SOT}. A $\langle sc \rangle$ symbol is inserted between the reference label sequences of speakers to represent speaker change. Furthermore, the reference label sequences are ordered according to their start time in a first-in first-out manner. In this way, the training loss is significantly simplified as
\begin{equation}
    \label{eq:sot}
    L^{SOT} = CE( {\bf{y}}, \Psi(1,...,S)).
\end{equation}

Given the large demand of streaming ASR in the industry, the backbone model of multi-talker ASR is also moving from the non-streaming AED model to the streaming RNN-T model. Although RNN-T was used in \cite{tripathi2020end} for multi-talker ASR, it's encoder is a bi-directional LSTM which is a non-streaming setting. In \cite{lu2021streaming, sklyar2021streaming}, the Streaming Unmixing and Recognition Transducer (SURT) was proposed as in Figure \ref{fig:HEAT}. A mask-based unmixing module is used to estimate masks in order to separate the speech input into two branches for recognition with RNN-T. Although PIT ASR loss in Eq. \eqref{eq:pitasr} can be used, it will introduce large computational cost due to the label permutation.  Therefore, Heuristic Error Assignment Training (HEAT) \cite{lu2021streaming} was proposed by ordering the label sequences based on the utterance start time into the set $\Omega(1,...,S)$. The loss function can be written as
\begin{equation}
    \label{eq:heat}
    L^{HEAT} = \sum_{s=1}^S CE( {\bf{y}}^s, {\bf{r}}^{\omega[s]}),
\end{equation}
where $\omega[s]$ stands for the $s$-th element of $\Omega(1,...,S)$. HEAT clearly reduces the training cost without losing accuracy \cite{lu2021streaming}. It is even more important to use Eq. \eqref{eq:heat} in the continuous streaming setup where it is formidable to have all the permutations in a long conversation \cite{raj2022continuous}.

\subsection{Multi-channel Models}
\begin{figure}[t]
  \centering
  \includegraphics[width=0.9\linewidth]{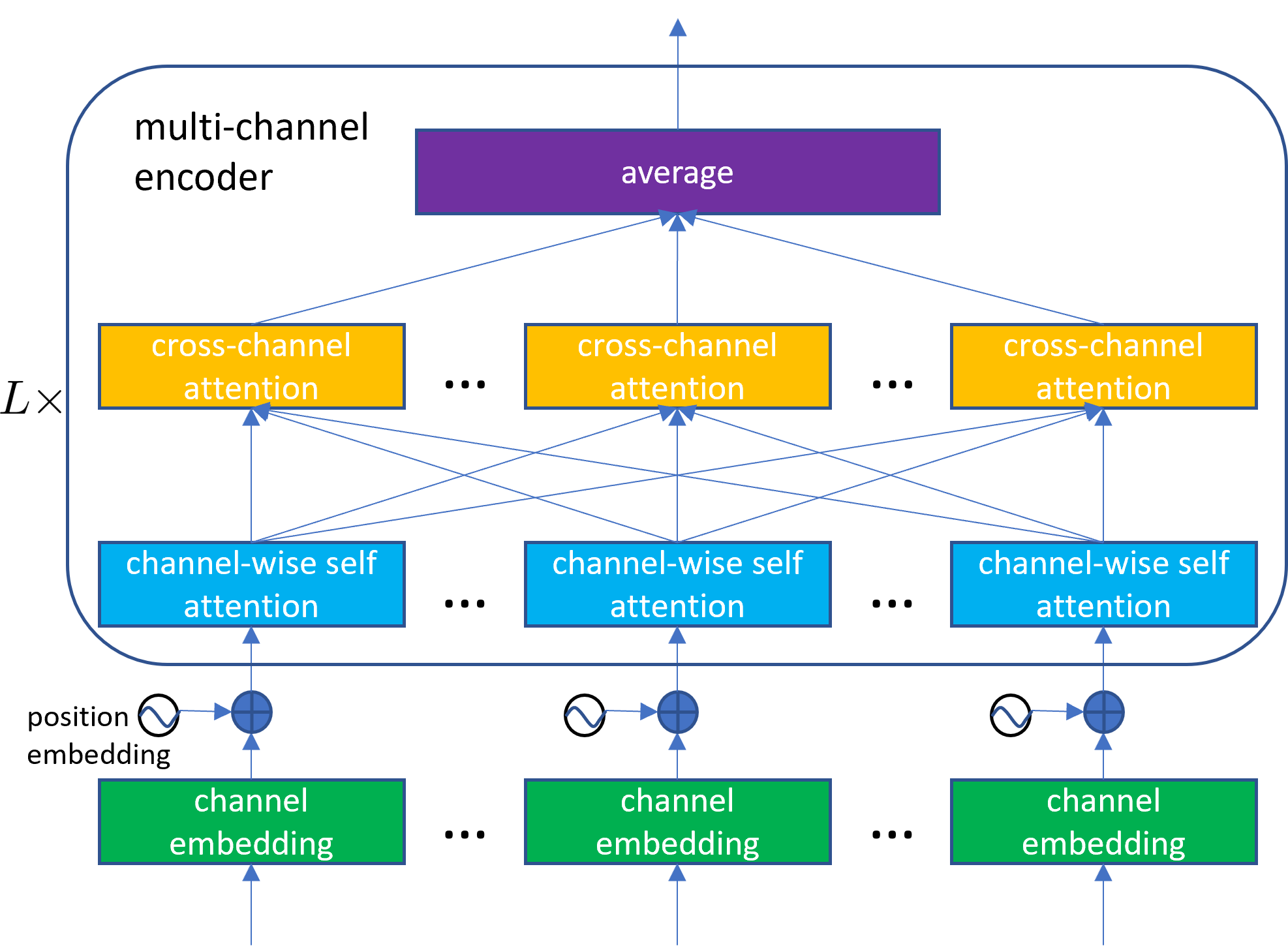}
  \caption{An example of multi-channel encoder \cite{chang2021end}}
  \label{fig:MCE}
\end{figure}

Beamforming is a standard technique for improving distant ASR system accuracy using  microphone arrays with multi-channel inputs \cite{omologo2001speech, kumatani2012microphone}. The most popular beamforming technology for ASR is signal processing based superdirective beamforming, while there is a trend to replace it with the neural beamforming for joint optimization with the backend hybrid models \cite{heymann2016neural, xiao2016deep, sainath2017multichannel, minhua2019frequency}. The joint optimization is more direct when the backend model is an E2E model \cite{ochiai2017multichannel, wang2021exploring}, and can be applied with even more front end components \cite{zhang2020end, zhang2021end}. While all these methods still work on good neural beamforming modules, some recent studies try to bypass the beamforming module design by using a single E2E network to preform ASR directly on multi-channel inputs. Thanks to the power and flexibility of Transformer, the multi-channel Transformer ASR works \cite{chang2021end, chang2021multi, gong2021self} replace the well-established beamformer with a multi-channel encoder which consumes the multi-channel inputs from microphone arrays. Figure \ref{fig:MCE} shows an example of a multi-channel encoder which consists of multiple blocks of cascaded within channel-wise self attention layer and cross-channel attention layer. The channel-wise self attention layer models the correlation across time within a channel while the cross-channel attention layer tries to learn the relationship across channels. The multi-channel encoder can then be plugged into any E2E model. Such E2E models do not need the conventional knowledge of beamformer design and were reported to perform  better than the standard beamforming method followed by an ASR model \cite{chang2021end, chang2021multi, gong2021self}.

\section{Miscellaneous Topics}

E2E modeling is a huge research topic. Due to the limit of space, we can only cover the most important areas we think about from the industry point of view. There are more areas worth mentioning. First of all, great toolkits \cite{doetsch2017returnn, watanabe2018espnet, wang2019espresso, shen2019lingvo, ravanelli2021speechbrain, yao2021wenet} are indispensable to the fast development of E2E ASR technologies. Among them, ESPnet \cite{watanabe2018espnet} is the most popular E2E toolkit widely used in the speech community although some companies have their own in-house E2E model training tools. We will continue to witness the growth of E2E toolkits which benefit the whole speech community.

Model unit selection is a very important area in E2E modeling. While most units (characters \cite{hannun2014deep, graves2014towards}, word-piece \cite{schuster2012japanese, chiu2018state, zenkel2018subword, xiao2018hybrid, huang2019exploring}, words \cite{soltau2016neural,  li2017acoustic, audhkhasi2018building}, or even phrases \cite{gaur2019acoustic})  are purely derived based on text and word-piece unit is the dominating one, there are also studies working on phoneme units \cite{sainath2018no, irie2019model, wang2020investigation, zhou2021phoneme}. As ASR is unique with the underlying acoustic signal, it should be better to build units bridging between the text and phoneme worlds. To that end, the pronunciation guided unit \cite{xu2019improving, zhou2021acoustic, papadourakis2021phonetically} is worth studying. 

Trained with paired speech-text data, E2E models are more data hungry than hybrid models. If the training data size is small, the performance of E2E models drops significantly. There are lots of ways to address the challenge of building E2E models for low-resource languages. First, regularization technologies such as dropout \cite{srivastava2014dropout} can be used to prevent E2E models from overfitting to limited training data \cite{zhou2017improved}. Data augmentation methods such as SpecAugment \cite{park2019specaugment} and speed perturbation \cite{ko2015audio} are also very helpful. In addition, there are also methods such as adversarial learning \cite{drexler2018combining} and meta-learning \cite{hsu2020meta}. 
The most popular solution is to first pre-train E2E models either with multilingual data or with self-supervised learning (SSL), and then fine-tune with the low-resource labeled data. A multilingual E2E model already captures lots of information across languages, which makes the transfer learning using the target language data very effective \cite{rosenberg2017end, cho2018multilingual, dalmia2018sequence, inaguma2019transfer, joshi2020transfer}. 
SSL is even more powerful because it does not need any labeled data for pre-training, naturally solving the low-resource challenge. Therefore,  SSL is becoming a new trend which especially works very well for ASR on resource limited languages \cite{schneider2019wav2vec, chung2019unsupervised, baevski2019vq, baevski2020wav2vec, chung2020generative, zhang2020pushing, wang2021unispeech, hsu2021hubert}, with representative technologies such as wav2vec 2.0 \cite{baevski2020wav2vec}, autoregressive predictive coding \cite{chung2019unsupervised}, and HuBERT \cite{hsu2021hubert}.  While most SSL studies focus on very limited supervised training data (e.g., 1000 hours), there are also recent studies showing promising results on industry-scale tens of thousand hours supervised training data \cite{wang2021unispeech2, zhang2021bigssl}.  

When training E2E ASR models, separate utterances are usually presented to the trainer. Therefore most ASR models are designed to recognize independent utterances. When the model is used to recognize a long conversation, a common practice is to segment the long conversation into utterances and then recognize them independently. However, the context information from previous utterances is useful to recognize the current utterance. In \cite{hori2020transformer, hori2021advanced, schwarz2021improving}, audio and decoded hypotheses from previous utterances are concatenated as the additional input to the model when processing the current utterance. 

When deploying E2E models to production, it is important to have an efficient decoding strategy, which was explored in \cite{saon2020alignment, prabhavalkar2021less}. Because the prediction network does not fully function as an LM as discussed in Section \ref{sec:adaptation}.\ref{ssec:domain_adapt}, the LSTM/Transformer in the RNN-T prediction network was recently replaced  with a simple and cheap embedding with very limited context, which can be used to significantly reduce decoding cost and model size \cite{botros2021tied, sainath2021efficient}. 
When deployed to small-footprint devices, model compression \cite{pang2018compression, mehrotra2020iterative, panchapagesan2021efficient},  quantization \cite{nguyen2020quantization, fasoli20214}, and the combination of multiple technologies \cite{shangguan2020optimizing} should be considered.  Confidence measure and word timing are sometimes required for practical E2E systems. For example, in a video transcription system, confidence score is used to indicate which word may be recognized wrongly and then the user can listen to the video corresponding to that word using its timestamp. Examples of confidence measure work for E2E systems are \cite{oneactua2021evaluation, li2021confidence, qiu2021learning, zhao2021addressing}, and examples of word timing  work for E2E systems are \cite{sainath2020emitting, zhao2021addressing}.

Almost all E2E models take Mel filter bank feature which is extracted from speech waveform as the input. In order to do the recognition really from end to end, speech waveform  as the input to E2E models was studied in \cite{collobert2016wav2letter, zeghidour2018end, zeghidour2018fully,  parcollet2020e2e, lam2021raw}, especially in recent influential wav2vec series work \cite{schneider2019wav2vec, baevski2020wav2vec}.
Because both E2E models and hybrid models have their own advantages and different error patterns, some works try to combine them together via rescoring \cite{li2019integrating}, minimum Bayes' risk combination \cite{wong2020combination}, or two-pass modeling \cite{ye2022have}. 
It is also very easy for E2E models to integrate inputs with multi-modality especially audio and visual signal together. There are plenty of works showing the benefits of visual signal for E2E ASR \cite{petridis2018end, afouras2018deep, petridis2018audio, sterpu2018attention, zhou2019modality, tao2020end, ghorbani2021listen, ma2021end}.

Robustness is always an important topic to ASR \cite{li2014overview}. E2E models tend to fit training data and should have even severe robustness challenge due to the mismatch between training and testing. However, there is not too much activity in addressing such mismatch \cite{andrusenko2020towards, prasad2021investigation}. T/S learning was used to adapt a clean-trained E2E model to a noisy environment \cite{meng2019domain}. Noise-invariant feature was learned to improve robustness in \cite{liang2018learning, zhang2020learning}. Data augmentation \cite{tsunoo2021data} is another effective way to expose more testing environments to E2E models during training.

\begin{table*}[t]
  \caption{Representative non-streaming E2E works on Librispeech}
  \label{tab:non-streaming}
  \centering
  \begin{tabular}{l|c|c|c|c}
    	\hline
			work with key technologies			& 	year & model & encoder & test-clean/other WER       \\							
    	\hline
        Deep Speech 2:  more labeled data, curriculum learning   \cite{amodei2016deep}     & 2016	&   CTC	&   bi-RNN	        	& 5.3/13.2  \\
    	policy learning, joint training \cite{zhou2018improving} & 2018 & CTC & CNN+bi-GRU & 5.4/14.7 \\
    	Shallow fusion, BPE, and pre-training  \cite{zeyer2018improved} & 2018	&   AED	&   BLSTM	    &   3.8/12.8  \\
    	ESPRESSO recipe: lookahead word LM, EOS thresholding   \cite{wang2019espresso}  & 2019	&   AED	&   CNN+BLSTM	&    2.8/8.7 \\
    	SpecAugment \cite{park2019specaugment}  & 2019	&   AED &	CNN+BLSTM	 &   2.5/5.8 \\
    	ESPnet recipe:  SpecAugment, dropout   \cite{karita2019comparative}    & 2019  & AED   &   Transformer     &   2.6/5.7 \\
    	Semantic mask \cite{wang2019semantic}       & 2019	&   AED	&   Transformer	&  2.1/4.9   \\
    	Transformer-T, SpecAugment \cite{zhang2020transformer} & 2020  &	RNN-T &	Transformer      &	2.0/4.6 \\
    	Conformer-T, SpecAugment   \cite{gulati2020conformer}  & 2020	&   RNN-T &	Conformer	&	1.9/3.9 \\
		wav2vec 2.0: SSL with unlabeled data,  DataAugment \cite{baevski2020wav2vec} & 2020  &	CTC &	Transformer & 1.8/3.3 \\
		internal LM prior correction, EOS modeling\cite{zeyer2021librispeech} & 2021  &   RNN-T   & BLSTM   & 2.2/5.6   \\
		w2v-BERT: SSL with unlabeled data,  SpecAugment \cite{chung2021w2v} & 2021  &	RNN-T &	Conformer   &	1.4/2.5 \\
		
    	\hline
  \end{tabular}
\end{table*}

\begin{table*}[t]
  \caption{Representative streaming E2E works on Librispeech}
  \label{tab:streaming}
  \centering
  \begin{tabular}{l|c|c|c|c}
    	\hline
			work with key technologies			& 	year & model & encoder &  test-clean/other WER        \\							
    	\hline
        stable MoChA, truncated CTC prefix probability \cite{miao2019online}  &   2019    & AED & LC-BLSTM & 6.0/16.7 \\
    	triggered attention  \cite{moritz2019streaming} & 2019  &	AED &   time-delayed LSTM	 &	5.9/16.8 \\
    	triggered attention, restricted self-attention, SpecAugment   \cite{moritz2020streaming} & 2020  &	AED &   Transformer	  &	2.8/7.2 \\
		Transformer-T,  restricted self-attention, SpecAugment \cite{zhang2020transformer} & 2020  &	RNN-T &	Transformer	&       2.7/6.6 \\
    	scout network, chunk  self-attention, SpecAugment \cite{wang2020low} & 2020  &	AED &	Transformer &	2.7/6.4 \\
		dual casual/non-casual self-attention, SpecAugment \cite{moritz2021dual} & 2021  &	AED &	Conformer     &	2.5/6.3 \\
		
    	\hline
  \end{tabular}
\end{table*}

Last but not least, it is always beneficial to examine technologies using a public database. There are lots of E2E works evaluated on Librispeech \cite{panayotov2015librispeech}, pushing the state-of-the-art (SOTA) forward. Table \ref{tab:non-streaming} and Table \ref{tab:streaming} give some representative works on Librispeech with non-streaming and streaming E2E models, respectively \footnote{Factors such as model size and latency are not listed in the Tables. They all affect the final word error rate (WER) of E2E models, especially streaming E2E models. Therefore, a method obtaining lower WER than another does not always mean that method is superior.}. It is amazing to see how significantly WER is reduced within only a few years, thanks to fast developing technologies.   There are two clear trends: from CTC to AED/RNN-T in the model aspect and from LSTM to Transformer in the encoder aspect. As achieving SOTA results on Librispeech is the goal of most works, there are much more non-streaming E2E studies. Because Librispeech only has 960 hours of labeled data, methods helpful to limited resource scenarios such as SSL and data augmentation are essential to boost the ASR accuracy on Librispeech.  

\section{Conclusions and Future Directions}

In this paper, we gave a detailed overview of E2E models and practical technologies that enable E2E models to not only surpass hybrid models in academic tasks but also potentially replace hybrid models in the industry. Although CTC is revived with advanced encoder structures, the most popular E2E models are AED and RNN-T. Because of the streaming nature, research has been gradually shifted from AED to RNN-T. The encoder is the most important module in E2E models, and attracts most research. The trend is also very clear -- the encoder structure has shifted from LSTM to Transformer and its variants.  Masking strategy is used to design efficient transformer encoders with low latency and small computational costs for industry application. In order to serve multilingual users well, multilingual E2E models are trained by pooling all the language data together. However, there is still a clear accuracy gap between the multilingual models with and without language ID. A configurable multilingual model can fill in the gap by training the model once and being configured based on the language selection by any multilingual user. Adaptation may be the most important area to work on in order to enable E2E models to replace hybrid models in the industry because of the huge accuracy improvement with adaptation for new speakers and new domains. Specifically, there is more research on domain adaptation and customization than speaker adaptation. The successful domain adaptation technologies usually have better use of text only data from the new domain while the successful customization methods can deal with the challenge of a very large context biasing list. In addition to the standard E2E model training criterion, T/S training is used to  either learn a small student model approaching the performance of a large teacher model or learn a streaming student model approaching the performance of a non-streaming teacher model. MWER training is to ensure E2E models are optimized with the training criterion consistent with ASR evaluation metrics.  Recently, there is a trend of developing advanced models such as non-autoregressive models which have much fast decoding speed by doing inference with one shot; unified models which are trained once but can be flexibly configured to meet different runtime requirements; and two-pass models which can leverage the advantages of both streaming and non-streaming models.  

A few years ago, there were very few people who believed E2E models would replace hybrid models which have lots of features designed for commercial ASR products. Within a very short time, we have witnessed that E2E models not only surpass hybrid models in terms of accuracy but also catch up with the practical features in commercial ASR systems. Because of the power of E2E modeling, the trend is not only  building  E2E models to replace traditional ASR models   but also   unifying speech processing modules such as speech separation, signal processing, and speaker identification into a single E2E ASR model. Multi-talker E2E models and multi-channel E2E models are examples of such a direction. The multi-talker E2E modeling was further extended with identifying speakers  \cite{kanda2020joint, lu2021streaming2}. There are also works of joint ASR and speaker diarization using E2E models by inserting speaker category symbols into ASR transcription  \cite{el2019joint, mao2020speech, soltau2021understanding}. As E2E ASR models directly map speech signals into target word sequences, they can be extended to E2E speech translation models when the target word sequence is from another language \cite{weiss2017sequence, di2019adapting, inaguma2019multilingual}. The major challenge for E2E speech translation is how to get enough training data with paired speech in the source language and text in the foreign language. The amount of such speech translation data is much smaller than the amount of ASR training data. Another challenge is how to handle the word reordering \cite{chuang2021investigating}. 

Although E2E modeling has already become the dominating ASR technology, there are still lots of challenges to be addressed before E2E models fully replace traditional hybrid models in both academic and industry. First, in hybrid modeling, paired speech-text data is used to build an acoustic model while  a very large amount of text only data is used to build LM. In contrast, general E2E models are only built with paired speech-text data. How to leverage the text only data to improve the accuracy of E2E models instead of simply doing LM fusion is a future direction to explore. In \cite{renduchintala2018multi}, a multi-modal data augmentation method was proposed by having two separate encoders: one for acoustic input and the other for text input, sharing the same attention and decoder modules. There are some cycle-consistency works using unpaired speech-text data to improve ASR accuracy \cite{tjandra2017listening, baskar2019semi}. Because the amount of text only data is much larger than that of the paired speech-text data, the cost is formidable to generate TTS audio from such large scale text only data for industry application although it is doable for small scale.  TTS audio sometimes also degrades the recognition accuracy on real speech \cite{li2019semi}. Therefore, although using TTS audio for domain adaptation is a good practice, it may not be a good way to build an industry-scale E2E model by synthesizing TTS audio from the text data used for LM training.  In \cite{sainath2020attention, mavandadi2021deliberation}, a joint acoustic and text decoder was proposed to leverage the text data by synthesizing TTS audio from the text to improve the decoder in AED. This was simplified in \cite{wang2021multitask} by  using multitask training with text only data. In \cite{bai2019learn, futami2020distilling}, an LM trained on large scale text is used as a teacher model to generate soft labels as the regularization of the AED model training. One promising way is the factorized neural transducer work \cite{chen2021factorized} in Figure \ref{fig:fnt} which has a standalone block working as a neural LM. Therefore, that LM-function block can be trained with large-scale text data efficiently. 

Another challenge is how to integrate knowledge into a single E2E model. For example, it is very easy for an E2E model with display format output to generate ``5:45'' when a user says ``five forty five''. However, it is very hard for an E2E model to output ``5:45'' when a user says ``a quarter to six'' because standard E2E models do not have the knowledge unless the training has seen such examples. 

In hybrid models, it is very easy to continuously add any new word not seen during training. However, this is very challenging for E2E models. The customization work described in Section \ref{sec:adaptation}.\ref{ssec:customization} provides a way to bias the ASR results towards a contextual phrase list. This is different from expanding the E2E models' capability of recognizing new words without biasing the ASR results. Although there are few works that allow injecting any out-of-vocabulary word into the vocabulary of acoustic-to-word E2E models during inference via subword-to-word embedding \cite{settle2019acoustically, collobert2020word}, this problem is far from being solved. 

Finally, it is always challenging to train good E2E models with low-resource languages. Given, the recent success of SSL which does not need any labeled data to pre-train a representation for downstream tasks, we predict that SSL will be closely coupled with E2E models in the near future. 

\section*{Acknowledgement}
The author would like to thank the anonymous reviewers, Xie Chen, Yashesh Gaur, Yanzhang He, Yan Huang, Naoyuki Kanda, Zhong Meng, Yangyang Shi, Eric Sun, Xiaoqiang Wang, Yu Wu, Guoli Ye, and Rui Zhao, for providing valuable inputs to improve the quality of this paper.

\newpage

\bibliographystyle{IEEEtran}

\bibliography{mybib}

\end{document}